\documentclass[prb,aps,twocolumn,amsmath,amssymb,floatfix,superscriptaddress]{revtex4}

\usepackage{amsmath}
\usepackage{amsfonts}
\usepackage{amssymb}
\usepackage{alphabeta}
\usepackage[dvips]{graphics}
\usepackage{color}
\definecolor{dred}{rgb}{0.75,0,0}
\usepackage{soul}
\usepackage{paralist}
\usepackage[colorlinks=true, citecolor=blue, urlcolor=blue ]{hyperref}
\date{\today}

\begin{document}
	
\title{Bias-driven circular currents in a quantum ring: Effects of electron-electron and electron-phonon interactions}

\author{Moumita Mondal}

\email{moumitamondal$_$r@isical.ac.in}

\affiliation{Physics and Applied Mathematics Unit, Indian Statistical Institute, 203 Barrackpore Trunk Road, Kolkata-700 108, India}

\author{Santanu K. Maiti}

\email{santanu.maiti@isical.ac.in}

\affiliation{Physics and Applied Mathematics Unit, Indian Statistical Institute, 203 Barrackpore Trunk Road, Kolkata-700 108, India}

\begin{abstract}
	
The phenomenon of bias-driven circular charge and spin currents in a ring nanojunction is investigated in the presence of electron-electron
(e-e) and electron-phonon (e-ph) interactions within a tight-binding framework based on the non-equilibrium Green's function formalism. The
Lang-Firsov transformation is employed to map the interacting system onto an effective electronic model, which is subsequently treated within
the Hartree-Fock mean-field scheme. By exploring the interplay among e-e interaction, e-ph coupling, and electrode-ring interface sensitivity,
several intriguing features emerge in both circular charge and spin currents that, to the best of our knowledge, have not been reported
previously. In addition to bias-driven circular currents, charge and spin-dependent junction currents through the nanojunction are also
analyzed. Selective spin transport is achieved, leading to a high degree of spin polarization. All four current components, two associated
with circular currents and two with transport currents, are systematically inspected over a wide range of parameter regimes to assess the
sensitivity of the results to the relevant system parameters. Our findings provide useful insights into charge and spin transport phenomena
in interacting nanojunctions with single- and multi-loop geometries.

\end{abstract}

\maketitle

\section{Introduction}

Electronic transport in the nanoscale regime has become a focal area of research, driven by its applications in spintronics, 
quantum computing, and related fields. In this regime, systems exhibiting loop geometries have attracted considerable
attention~\cite{bs1,bs2,bs3,bs4,bs5}. A non-dissipative loop current induced by a magnetic flux piercing the loop, commonly 
referred to as a flux-driven persistent current, was theoretically predicted and experimentally verified several years
ago~\cite{pc1,pc2,pc3,pc4,pc5,pc6}. More recently, considerable attention has been devoted to the possibility of
driving a current circulating within a conducting ring by an external voltage bias rather than by a magnetic 
flux~\cite{cir1,cir2,cir3,cir4,cir5}. Such a bias-driven circular current can, in turn, generate a local magnetic field ranging 
from a few milliTesla to a few Tesla. In these open quantum systems, the magnitude of the transport (junction) 
current~\cite{jn1,jn2,jn3,jn4}, which characterizes the overall conduction through an electrode-conductor nanojunction,
can sometimes be smaller than the magnitude of the current circulating within the loop.

Compared to flux-driven persistent currents in `isolated' quantum loops~\cite{pc1,pc2,pc3,pc4,pc5,pc6}, which are not attached to 
external electronic baths, the phenomenon of bias-driven circular currents in ring-like geometries coupled to external electrodes, 
referred to as open quantum systems, is relatively new, and a limited amount of work has been carried out so
far~\cite{cir1,cir2,cir3,cir4,cir5}. To the best of our knowledge, existing studies of bias-driven circular currents have largely 
focused on non-interacting electrons. This is, of course, a reasonable approximation. However, for a more complete description, 
the consideration of electron-electron (e-e) interaction and electron-phonon (e-ph) coupling is highly relevant.
The e-ph interaction plays a significant role in determining the transport characteristics of various systems.

To investigate the roles played by e-e and e-ph interactions in bias-driven circular currents, we consider a one-dimensional (1D)
non-magnetic quantum ring in which both interactions are present. The e-e interaction is incorporated within the ring following the
1D on-site Hubbard model, where two electrons of opposite spin occupying the same atomic site interact through Coulomb 
repulsion~\cite{ee1,ee2,ee3,ee4,ee5,ee6,ee7}. The e-ph interaction is introduced through the Holstein model, which accounts for the 
interaction of itinerant electrons with phonons~\cite{ep1,ep2,ep3,ep4,ep5}. 
We consider dispersionless optical phonons that vibrate out of plane, following Einstein's model, in which all phonons possess the 
same energy. The effect of acoustic phonon modes is neglected because their mean free paths are much longer than those of the
optical modes, resulting in a comparatively negligible contribution to electron scattering~\cite{exep1}. Each of these interactions, 
e-e and e-ph,
gives rise to rich transport physics, and their interplay can lead to further interesting behavior, which is the focus of the present 
study. The inclusion of e-e interaction develops magnetization at different lattice sites, corresponding to local magnetic moments, and the
interaction of itinerant electrons with these local magnetic moments leads to spin-selective electron transport. The effect becomes more
pronounced when e-ph coupling is included, as it has two major effects. An indirect e-e interaction is induced, which modifies the
existing Hubbard interaction strength, while the effective electronic mass is enhanced, thereby significantly suppressing the bandwidth.

Along with the bias-driven charge and spin circular currents, the transport (junction) charge and spin currents are also inspected in
our ring-electrode junction setup. From the spin-resolved currents, we also evaluate the spin polarization coefficient. The inclusion
of e-ph coupling leads to an overall suppression of the transport current, which is directly related to the enhancement of the effective
mass of electrons, whereas the behavior of the circular current is more complex, as it depends on several factors, including the degree
of asymmetry between the clockwise- and anticlockwise-propagating electronic waves in the different arms of the loop, the choice of the
Fermi energy, and the electrode-ring interface configurations.

We simulate the quantum system within a tight-binding (TB) framework with nearest-neighbor electron hopping and incorporate the
interactions following the well-known Hubbard-Holstein (HH) model~\cite{ep1,ep2,ep5}. The HH model provides one of the simplest 
frameworks for treating these interactions. The e-ph-coupled Hamiltonian is renormalized into an effective electronic Hamiltonian 
using the Lang-Firsov (LF) unitary transformation~\cite{ep7,ep8}. The resulting effective interacting electronic system is further 
treated within the Hartree-Fock (HF) mean-field (MF) scheme using a self-consistent approach~\cite{mf1,mf2,mf3,mf4}. 
Finally, all four current components, namely the charge and spin currents in the ring and in the 
drain, are evaluated following the Green's function method~\cite{bs3,bs4,bs5,gf1}. The key aspects that we aim to explore through 
our detailed numerical results 
are: (i) generation and manipulation of spin currents, and hence spin polarization, through e-e interaction, e-ph coupling, or both, 
(ii) enhancement of bias-driven circular currents by tuning e-ph coupling~\cite{exep2,exep3,exep4,exep5,exep6}, (iii) achievement of 
a high degree of spin polarization by
controlling e-ph and e-e interactions, and (iv) exploration of the critical role of the ring-electrode junction configuration on the 
different current components.

The rest part of the work is organized as follows. In Sec. II, we describe the ring-electrode junction setup, the TB Hamiltonian, and the
mathematical tools used for obtaining the results. Section III presents and discusses the numerical results in detail. The essential 
findings are summarized in Sec. IV. Some relevant mathematical details and derivations are provided in different appendices for 
completeness of our study.

\section{Ring nanojunction, TB Hamiltonian, and theoretical framework}

This section is divided into two sub-sections. In the first sub-section, we describe the model ring nanojunction and the corresponding
TB Hamiltonian. In the second sub-section, we outline the theoretical framework and mathematical methods used to calculate the various
quantities required for our analysis.

\subsection{Ring nanojunction and TB Hamiltonian}

\begin{figure}[ht]
\centering \resizebox*{7.5cm}{4.6cm}{\includegraphics{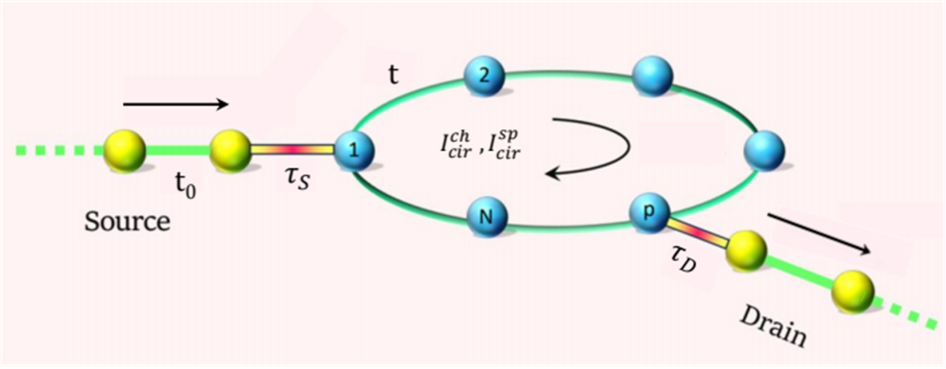}}
\caption{(Color online). Schematic view of the junction setup, where a quantum ring is coupled to source and drain electrodes. Along with
transport (junction) currents, circular currents are obtained when a bias voltage is applied across the electrodes.}
\label{fig1}
\end{figure}

Let us begin with the schematic diagram shown in Fig.~\ref{fig1}, where an $N$-site quantum ring (with even $N$) is coupled to two
one-dimensional electrodes, referred to as the source (S) and the drain (D). When a bias voltage ($V$) is applied across these electrodes,
a circulating current may emerge in the ring. If $I_1$ and $I_2$ denote the currents flowing through the two arms of the ring, the
circulating current is defined as $I_{cir} = (I_1 L_1 + I_2 L_2)/L$, where $L_1$ and $L_2$ are the lengths of the two arms, and
$L = L_1 + L_2$ ($=Na$, $a$ being the lattice spacing) is the circumference of the ring. For a symmetric junction configuration, 
$L_1 = L_2$ and $I_1 = -I_2$, which results
in $I_{cir} = 0$. Therefore, to generate a non-zero circulating current, the symmetry of the system must be broken. This can be achieved
in two distinct ways: (i) by asymmetrically connecting the electrodes to an otherwise perfect ring, such that the two arms have different
lengths ($L_1 \neq L_2$), and (ii) by introducing different physical conditions in the two arms while maintaining a lengthwise symmetric
configuration~\cite{cr1,cr2,cr3,cr4,cr5}.

The full Hamiltonian of the junction setup can be divided into three parts: the semi-infinite electrodes ($H_{S/D}$), the conducting 
ring ($H_R$), and the coupling between the ring and the electrodes ($H_{coupling}$). We describe these Hamiltonians one by one as follows.

The TB Hamiltonian for the source (drain), including the spin degrees of freedom, can be written as
\begin{equation}
H_{S/D}=\sum_{i,\sigma} \epsilon{^0_{i\sigma}} d{_{i\sigma}^\dagger}d_{i\sigma}  + t_0\sum_{<i,j>,\sigma} \left[d{_{i\sigma}^{\dagger}} d_{j\sigma} + d{_{j\sigma}^{\dagger}} d_{i\sigma}\right]
\label{eqn1}
\end{equation}
where $d{_{i\sigma}^\dagger}$ ($d_{i\sigma}$) is the electronic creation (annihilation) operator for an electron with spin $\sigma$
($\uparrow,\downarrow$) at the $i$th site, $\epsilon{^0_{i\sigma}}$ denotes the onsite energy, and $t{_0}$ is the nearest-neighbor 
hopping strength in the electrodes.

The quantum ring, in which electron-electron and electron-phonon interactions are present, can be described by the 1D
Hubbard-Holstein Hamiltonian~\cite{ep1,ep2},
\begin{eqnarray}
H_R&=&\sum_{i,\sigma} \epsilon_{i\sigma} n_{i\sigma}  + t\sum_{<i,j>,\sigma} \left[c{_{i\sigma}^{\dagger}} c_{j\sigma} + c{_{j\sigma}^{\dagger}} c_{i\sigma}\right]\nonumber\\
& + & u\sum_{i} n_{i\uparrow} n_{i\downarrow} + \hbar \omega_{0} \sum_i b{{_i}{^\dagger}}b_i \nonumber\\ 
& + & g\sum_{i,\sigma} (b{{_i}{^\dagger}}+ b_i) n_{i\sigma}.
\label{eqn2}
\end{eqnarray}
Here, $c{_{i\sigma}^\dagger}$ and $c_{i\sigma}$ are the usual fermionic creation and annihilation operators, respectively. 
$t$ is the hopping strength between nearest-neighbor sites. 
$n_{i\sigma} = c{_{i\sigma}^\dagger}c_{i\sigma}$ is the electron number operator. The Hubbard interaction strength is denoted by $u$, 
which is non-zero when a lattice site is doubly occupied. $\hbar \omega_0$ is the energy of the out-of-plane phonon modes. 
$b{_i^\dagger}$ ($b_i$) is the bosonic creation (annihilation) operator for these phonons, and $g$ is the e-ph coupling parameter.

The ring-electrode coupling Hamiltonian can be written as
\begin{equation}
H_{coupling}=\sum_\sigma (\tau_S c_{1\sigma}^\dagger d_{0\sigma}+\tau_D c_{p\sigma}^\dagger d_{N+1\sigma})+h.c.
\label{eqn3}
\end{equation}
where $\tau_S$ is the coupling strength between the source and the ring, and $\tau_D$ denotes the coupling strength between the ring and
the drain. We assume that the source is coupled to site $1$ of the ring, while the drain is connected to site $p$ (which is variable) of
the ring (see Fig.~\ref{fig1}). In our representation, the 1st sites of S and D with which the ring is coupled are labeled as `0' and 
($N+1$), respectively.

\subsection{Theoretical formulation}

\textbf{Lang-Firsov transformation}: Since the ring Hamiltonian (Eq.~\ref{eqn2}) contains both the e-e and e-ph interactions, we first 
employ the Lang-Firsov unitary transformation to eliminate the explicit e-ph coupling and obtain an effective electronic 
Hamiltonian~\cite{ep7,ep8}.
For that, we use a transformation generator expressed as 
\begin{equation}
U=\frac {g} {\hbar\omega_0} \sum_{i,\sigma} (b{{_i}{^\dagger}}-b_i) n_{i\sigma}
\label{eqn4}
\end{equation}
which is anti-Hermitian. It transforms the ring Hamiltonian as (detailed derivation is given in Appendix~\ref{ap1})
\begin{eqnarray}
\widetilde{H}_R &=& e^U\,  H_R \, e^{-U}\nonumber\\
& = & \sum_{i,\sigma}\left(\epsilon_{i\sigma}-\frac{g^2}{\hbar\omega_0}\right)c{_{i\sigma}^{\dagger}}c_{i\sigma}\nonumber\\
& +  &t \sum_{<i,j>,\sigma} \bigg\{c{_{i\sigma}^{\dagger}}c_{j\sigma}
\exp\left[\left(\frac{g}{\hbar\omega_0}\right)\left(b{_i^{\dagger}}-b_i\right)\right]\nonumber\\
&   &\exp\left[-\left(\frac{g}{\hbar\omega_0}\right)\left(b{_j^{\dagger}}-b_j\right)\right]+h.c\bigg\}\nonumber\\
& + & \left(u-\frac{2g^2}{\hbar\omega_0}\right) \sum_{i}n_{i\uparrow}n_{i\downarrow}+\hbar\omega_0\sum_i
b{_i^{\dagger}}b_i.
\label{eqn5}
\end{eqnarray}
We make an ansatz that the ground state of the Hubbard-Holstein Hamiltonian is given by 
$|\Psi\rangle=|\Psi_{el}\rangle \otimes e^U |0_{ph}\rangle$, where $|\Psi_{el}\rangle$ is the electronic wave function and 
$|0_{ph}\rangle$ is the phonon vacuum state. At absolute zero temperature, the zero-phonon averaging leads to the effective ring 
Hamiltonian as (see, Appendix~\ref{ap2} for the detailed derivation)
\begin{eqnarray}
H{_R^{eff}} & = & \left\langle 0_{ph}|\widetilde{H}_R|0_{ph}\right\rangle \nonumber \\
& = & \sum_{i,\sigma}\epsilon_{i\sigma}^{eff} n_{i\sigma} + t^{eff}\sum_{<i,j>,\sigma} \left[c{_{i\sigma}^{\dagger}} c_{j\sigma} 
+ c{_{j\sigma}^{\dagger}} c_{i\sigma}\right] \nonumber \\
& + & u^{eff}\sum_{i} n_{i\uparrow} n_{i\downarrow}
\label{eqn6}
\end{eqnarray}
where, the site energy, hopping strength, and Hubbard interaction strength are renormalized, and they are given by
\begin{eqnarray}
\epsilon_{i\sigma}^{eff} & = &\left(\epsilon_{i\sigma}-\frac{g^2}{\hbar \omega_0}\right), \nonumber\\
t^{eff}& = & t\exp\left[-\left(\frac{g}{\hbar \omega_0}\right)^2\right], \nonumber\\
u^{eff} & = & \left(u-\frac{2g^2}{\hbar \omega_0}\right). \nonumber
\end{eqnarray}
In a similar way, the effective coupling Hamiltonian can be obtained as
\begin{equation}
H{^{eff}_{coupling}}= \sum_\sigma e^{-\frac{1}{2}\left(\frac{g}{\hbar \omega_0}\right)^2} (\tau_S c_{1\sigma}^\dagger d_{0\sigma}+\tau_D c_{p\sigma}^\dagger d_{N+1\sigma})+h.c.
\label{eqn7}
\end{equation}
The effective many-body Hamiltonian in Eq.~\ref{eqn6} is treated within the mean-field approximation, where it can be
expressed in terms of the up and down spin electrons separately.

\vskip 0.2cm
\textbf{Mean-field scheme}: Under the mean-field scheme, the site energies for up- and down-spin electrons are modified 
as~\cite{mf1,mf2,mf3,mf4}
\begin{eqnarray}
\epsilon_{i\uparrow}^\prime=\epsilon{^{eff}_{i\uparrow}} + u^{eff} \langle n_{i\downarrow} \rangle \nonumber \\
\epsilon_{i\downarrow}^\prime=\epsilon{^{eff}_{i\downarrow}} + u^{eff} \langle n_{i\uparrow} \rangle \nonumber
\end{eqnarray}
and the ring Hamiltonian can be written as
\begin{eqnarray}
H^{eff}_R=H_{R,\uparrow}^{eff}+H_{R,\downarrow}^{eff}-u^{eff} \sum_{i} \langle n_{i\uparrow} \rangle \langle n_{i\downarrow} \rangle
\label{eqn8}
\end{eqnarray}
where,
\begin{eqnarray}
H_{R,\uparrow}^{eff} = \sum_{i} \epsilon_{i\uparrow}^\prime n_{i\uparrow} + t^{eff} \sum_{<i,j>} \left[c{_{i\uparrow}^{\dagger}}c_{j\uparrow} 
+ c{_{j\uparrow}^{\dagger}} c_{i\uparrow}\right],\nonumber \\
H_{R,\downarrow}^{eff} = \sum_{i} \epsilon_{i\downarrow}^\prime n_{i\downarrow} + t^{eff} \sum_{<i,j>} \left[c{_{i\downarrow}^{\dagger}}c_{j\downarrow} 
+ c{_{j\downarrow}^{\dagger}} c_{i\downarrow}\right]. \nonumber
\end{eqnarray}
To get the converged up and down spin sub-Hamiltonians, we start the self-consistent procedure with initial guess values of 
$\langle n_{i\uparrow} \rangle$ and $\langle n_{i\downarrow} \rangle$. These initial guess values are chosen based on the electron 
filling. The up and down spin sub-Hamiltonians are then constructed using these initial values. By diagonalizing $H_{R,\uparrow}^{eff}$
and $H_{R,\uparrow}^{eff}$, we obtain the eigenvalues and a new set of $\langle n_{i\uparrow} \rangle$ and 
$\langle n_{i\downarrow} \rangle$. The iteration is continued until the convergence is achieved.

\vskip 0.2cm
\textbf{Non-equilibrium Green's function (NEGF) approach}: The transmission probabilities, transport (junction) currents, and circular 
currents are obtained using the well-known NEGF formalism~\cite{bs3,bs4,gf1}. We need to define retarded, advanced, and correlated 
Green's functions, where the effects of side-coupled contacts are incorporated through self-energies. 
 
\vskip 0.2cm
\noindent 
$\blacksquare$ {\em Non-interacting system}: In such a case, the Green's functions are defined as
\begin{eqnarray}
G{_\sigma^r} & = & \left[E\,\mathbb{I}-H_{R,\sigma}^{eff}-\Sigma_S-\Sigma_D\right]^{-1} \nonumber\\
G{_\sigma^a} & = & \left(G{_\sigma^r}\right)^{\dagger}
\label{eqn9}
\end{eqnarray}
where, $G_\sigma^r$ and $G_\sigma^a$ are the retarded and advanced Green's functions, respectively. $E$ is the electronic energy and $I$
is the identity matrix. $\Sigma_S$ and $\Sigma_D$ are the contact self-energies of the source and drain, respectively, and they are obtained
from the relation $\Sigma_{S/D}=H_{coupling}\, g{^r_{S/D}}\, H_{coupling}^{\dagger}$. $g^r_S$ and $g^r_D$ are the retarded surface Green's 
functions of the source and drain, respectively. The contact self-energies ($\Sigma_S$, $\Sigma_D$) are determined entirely by the 
properties of the corresponding isolated lead and its coupling to the central region, i.e., the ring. The transmission probability is  
obtained from the relation
\begin{equation}
T_{\sigma}(E)=\mbox{Tr}\left[\Gamma_S G_{\sigma}^r \Gamma_D G_{\sigma}^a\right]
\label{eqn10}
\end{equation}
where $\Gamma_S=-2\,\mbox{Im}\left[\Sigma_S\right]$ and $\Gamma_D=-2\,\mbox{Im}\left[\Sigma_D\right]$ are the coupling matrices 
associated with S and D, respectively.

\vskip 0.2cm
\noindent 
$\blacksquare$ {\em Interacting system}: For the interacting case, the retarded and advanced Green's functions are defined as 
\begin{equation}
G_\sigma^r = \left(G{_\sigma^a}\right)^{\dagger}= \left[E\,\mathbb{I}-H^{eff}_{R,\sigma}-\Sigma{^{eff}_S}-\Sigma{^{eff}_D}\right]^{-1}
\label{eqn11}
\end{equation}
where the effective self-energies of S and D are obtained from the modified effective coupling Hamiltonian (see Appendix~\ref{ap3} for 
detailed calculation), and they are written as  
\begin{eqnarray}
\Sigma{_{S/D}^{eff}} & = & H{^{eff}_{coupling}}\,\,g{^r_{S/D}}\,\,(H{^{eff}_{coupling}})^{\dagger} \nonumber \\
& = & \exp\left[-\frac{1} {2} \left(\frac{g}{\hbar \omega_0}\right)^2\right]  H_{coupling}\,\,g{^r_{S/D}} \nonumber \\
&& \exp\left[-\frac{1} {2} \left(\frac{g}{\hbar \omega_0}\right)^2\right] 
(H_{coupling})^{\dagger}\nonumber \\
& = & \exp\left[-\left(\frac{g}{\hbar \omega_0}\right)^2\right] H_{coupling}\,\,g{^r_{S/D}}\,\,(H_{coupling})^{\dagger} \nonumber \\
& = & \exp\left[-\left(\frac{g}{\hbar \omega_0}\right)^2\right]\Sigma_{S/D}.
\label{eqn12}
\end{eqnarray}
The quantity $g{^r_{S/D}}$ remains unchanged for the interacting system because it contains information only about the isolated leads 
and is independent of the properties of the bridging conductor between S and D.

Due to the presence of e-ph interaction, the formula for the transmission probability gets modified and can be written as 
\begin{equation}
T_{\sigma}(E)=\mbox{Tr}\left[\Gamma_S G{_\sigma^r}\Gamma_D \Lambda G{_\sigma^a}\right].
\label{eqn13}
\end{equation}
In the linear-response regime, $\Lambda$ can be expressed as
\begin{equation}
\Lambda =\mathbb{I} \exp\left[-\left(\frac{g}{\hbar \omega_0}\right)^2\right]\nonumber
\end{equation}
One can use $\Lambda\sim\mathbb{I}$, in the limit of weak interaction.

The transport (junction) current is computed using the Landauer-B\"{u}ttiker formula
\begin{equation}
I{^\sigma_{tr}}=\frac{e}{h} \int_{-\infty}^\infty T_\sigma(E) \left[f_S(E)-f_D(E)\right]\,dE
\label{eqn14}
\end{equation}
where $$f_{S/D}(E)=\left[1+e^{\frac{\left(E-\mu_{S/D}\right)}{K_B\mathcal{T}}}\right]^{-1}.$$ 
$f_S$ and $f_D$ are the Fermi-Dirac distribution functions of S and D, respectively, and $\mu_S$ and $\mu_D$ are the corresponding 
electrochemical potentials which are defined as: $\mu_S=E_F+eV/2$ and $\mu_D=E_F-eV/2$. $E_F$ is the equilibrium Fermi energy, and
$\mathcal{T}$ is the absolute temperature. 

At absolute zero temperature, for a finite bias $V$, electrons can transmit through the energy window extending from $E_F-eV/2$ to
$E_F+eV/2$, and therefore,
\begin{equation}
I{^\sigma_{tr}}=\frac{e}{h} \int_{E_F-eV/2}^{E_F+eV/2} T_\sigma(E)\, dE.
\label{eqn15}
\end{equation}
From the spin-dependent current components, the total charge and spin transport currents are defined as
\begin{eqnarray}
I{_{tr}^{ch}}=I{_{tr}^{\uparrow}}+I{_{tr}^{\downarrow}}\nonumber \\
I{_{tr}^{sp}}=I{_{tr}^{\uparrow}}-I{_{tr}^{\downarrow}}
\label{eqn16}
\end{eqnarray}
where $I_{tr}^{\uparrow}$ and $I_{tr}^{\downarrow}$ are the transport currents carried by up and down spin electrons, respectively.
The spin polarization (SP) coefficient is defined as
\begin{equation}
SP=\frac {I_{tr}^{\uparrow} - I_{tr}^{\downarrow}}{I_{tr}^{\uparrow} + I_{tr}^{\downarrow}}.
\label{eqn17}
\end{equation}

To determine the bias-driven circular charge and spin currents, we first evaluate the bond current densities along the individual bonds 
of the ring using the correlated Green's function, which is defined as
\begin{equation}
G_\sigma^n= G_\sigma^r \, f_D\, \Gamma_D\, G_\sigma^a.
\label{eqn18}
\end{equation}
For an electron with spin $\sigma$ transmitted from site $i$ to site $j$, the bond current density is given 
by~\cite{bon1,bon2,bon3,bon4}
\begin{equation}
J^{\sigma}_{ij}=\left(\frac{2e}{h}\right) \mbox{Im}\left[\left(H_R^{eff}\right)_{ij}\left(G_\sigma^n\right)_{ij}\right]
\label{eqn19}
\end{equation}
At absolute zero temperature, the current corresponding to each bond is evaluated through the following Landauer-like expression:
\begin{equation}
I{_{ij}^\sigma}=\int_{E_F-eV/2}^{E_F+eV/2} J{_{ij}^\sigma}(E)\, dE.
\label{eqn20}
\end{equation}
Summing the contributions from all bonds, we obtain the net circular current for spin $\sigma$ as
\begin{equation}
I_{cir}^\sigma=\frac{1}{N} \sum_{<ij>} I{_{ij}^\sigma}
\label{eqn21}
\end{equation}
where $N$ is the total number of atomic sites in the quantum ring and appears in the denominator of the above relation following 
the usual definition of the bias induced circular current.
\begin{figure*}[ht]
\centering \resizebox*{14cm}{13cm}{\includegraphics{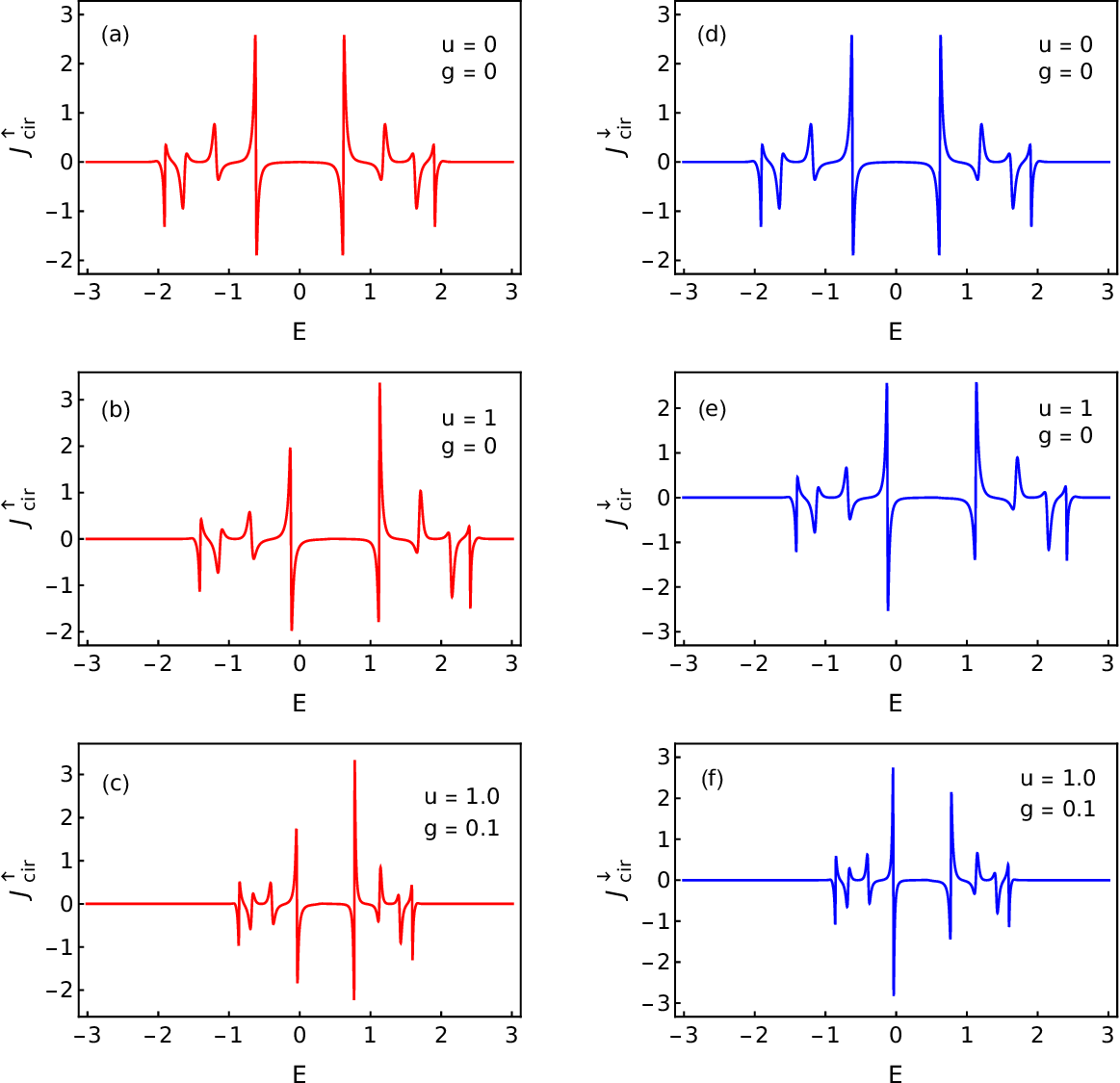}}
\caption{(Color online). (a)-(c) Circular current density ($J{^{ \uparrow}_{cir}}$) for spin-up electrons as a function of energy $E$ 
for different combinations of e-e interaction strength $u$ and e-ph coupling strength $g$. (d)-(f) Same as (a)-(c), but for 
spin-down electrons.}
\label{fig2}
\end{figure*}

Using the spin-dependent components, the charge and spin circular currents can then be written as
\begin{eqnarray}
I{_{cir}^{ch}}=I{_{cir}^{\uparrow}}+ I{_{cir}^{\downarrow}}\nonumber \\
I{_{cir}^{sp}}=I{_{cir}^{\uparrow}}- I{_{cir}^{\downarrow}}
\label{eqn22}
\end{eqnarray}
where $I_{cir}^{\uparrow}$ and $I_{cir}^{\downarrow}$ are the circular currents carried by the up and down spin electrons, respectively.

\section{Numerical Results and Discussion}

In accordance with the theoretical framework outlined above (Sec. II), this section presents and discusses the numerical results. 
Before proceeding to a detailed analysis, we first specify the parameter values used throughout the study. 
\begin{figure*}[ht]
\centering \resizebox*{14cm}{13cm}{\includegraphics{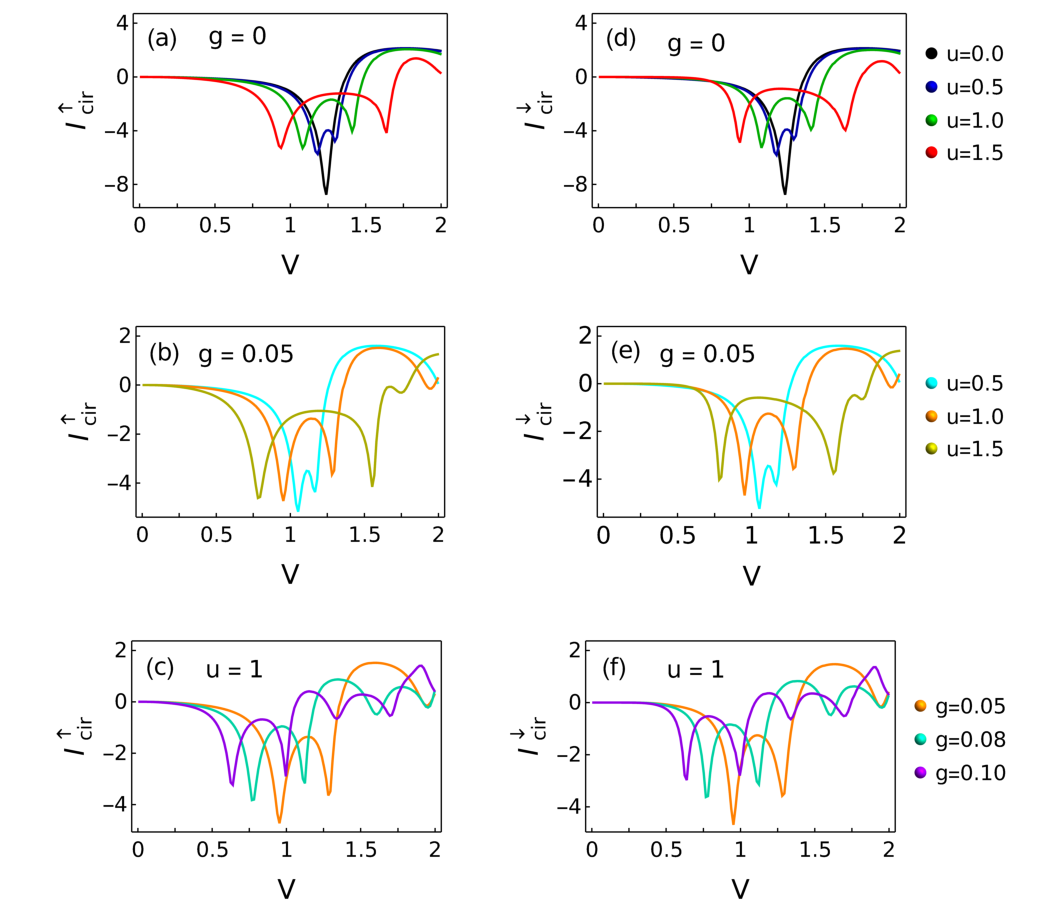}}
\caption{(Color online). Up and down spin circular currents, $I^{\uparrow}_{cir}$, $I^{\downarrow}_{cir}$, as a function of bias voltage
$V$, for some typical sets of $u$ and $g$. The left and right columns correspond to up and down spin electrons, respectively.}
\label{fig3}
\end{figure*}
For the semi-infinite source
and drain leads, $\epsilon_{i\uparrow}^0=\epsilon_{i\downarrow}^0=0$ and $t_0=2$. For the conducting ring, we set $t=1$ and $\epsilon_{i\uparrow}=\epsilon_{i\downarrow}=0$ for all $i$, where $\epsilon_{i\uparrow}$ ($\epsilon_{i\downarrow}$) is the site 
potential experienced by an up (down) spin electron at the $i$th site of the ring. The coupling strengths between the ring and the 
two leads are taken as $\tau_S=\tau_D=1$. The vibrational energy of all the phonons is taken to be $\hbar\omega_0=0.15$. We consider 
a $20$-site ring as the conductor, with each site capable of accommodating two electrons of opposite spin. These electrons interact 
through the Coulomb interaction with strength $u$. The ring is assumed to be half-filled, i.e., the total number of electrons is $20$, 
with $10$ up spin and $10$ down spin electrons. At half-filling, the ground state of the Hubbard model exhibits antiferromagnetic 
ordering, in which double occupancy is energetically disfavored and the neighboring magnetic moments tend to align along opposite 
quantized directions, $+Z$ and $-Z$. Unless otherwise specified, the source and drain are connected to the $1$st and $9$th sites of 
the ring, respectively. We also consider an alternative junction configuration at the end of this section to inspect the consistency 
of the results obtained. All results are evaluated at absolute zero temperature, and the equilibrium Fermi energy is set to the energy 
of the highest occupied molecular level of the conducting ring. For each combination of the parameters $u$ and $g$, a new set of 
eigenvalues is obtained, consequently, the Fermi energy is adjusted accordingly, and the corresponding current is evaluated within 
the associated Fermi window. The values of the other parameters are specified at the appropriate locations. All energies are measured 
in units of electron-volt (eV), and currents are expressed in units of $\mu$A.

Let us begin analyzing the results. When the energy of an incident electron resonates with an eigenvalue of the ring, a transmission 
peak generally occurs. Thus, the transmission peaks are expected to lie close to the eigenvalues of the isolated ring. Figure~\ref{fig2} 
shows the circular current density as a function of energy $E$ for up ($J{^{\uparrow}_{cir}}$) and down ($J^{\downarrow}_{cir}$) spin 
electrons. The peaks followed by dips in the circular current density profiles generally occur around those eigenvalues at which a 
transmission peak is present, and the corresponding eigenvalue is two-fold degenerate. These degenerate eigenvalues correspond to orbital
momenta of $+k$ and $-k$, representing clockwise- and anticlockwise-propagating Bloch waves, respectively. The coupling between the leads 
and the ring breaks the symmetry between these two Bloch waves when the upper and lower arms of the ring are asymmetric, either in length 
or in their physical properties, or in both.

Figures~\ref{fig2}(a) and (d) show the circular current density profiles for a non-interacting perfect ring connected to the source and 
drain. Figures~\ref{fig2}(b) and (e) show the corresponding results when the e-e interaction term is included. In these two panels, we 
observe an overall shift of the entire profile upon introducing the e-e interaction. This feature can be understood from the fact that, 
at half-filling, the Hubbard model favors an antiferromagnetic ground state, and the resulting up- and down-spin Hamiltonians resemble 
those of a bipartite lattice, leading to the opening of a HOMO-LUMO gap near the band center. Consequently, the positions of the minima 
and maxima of the circular current density are modified compared with the non-interacting case, in which all the site energies are identical.
Another notable feature is that, unlike the non-interacting case, the up (Fig.~\ref{fig2}(b)) and down spin (Fig.~\ref{fig2}(e)) circular
current density profiles are no longer identical, giving rise to spin-selective transmission. The separation between the up and down spin
eigenvalue spectra arises from the system's sensitivity to the junction configuration. In the presence of the Hubbard interaction, the 
up and down spin Hamiltonians acquire an effective bipartite-lattice structure, with a relative shift between the two sublattices, 
$A\rightarrow B$. The sublattice symmetry is broken by the connected leads for a suitable junction configuration. For the interacting 
nano-ring in our chosen junction configuration, the side-attached leads modify the density of states (DOS) of the entire system, resulting 
in slightly different DOS profiles for up- and down-spin electrons. In Figs.~\ref{fig2}(c) and (f), the presence of a non-zero e-ph coupling
leads to bandwidth narrowing and also renormalizes the e-e interaction, yielding an effective interaction strength $u^{eff}$ such that
\begin{figure}[ht]
\centering \resizebox*{7cm}{11.5cm}{\includegraphics{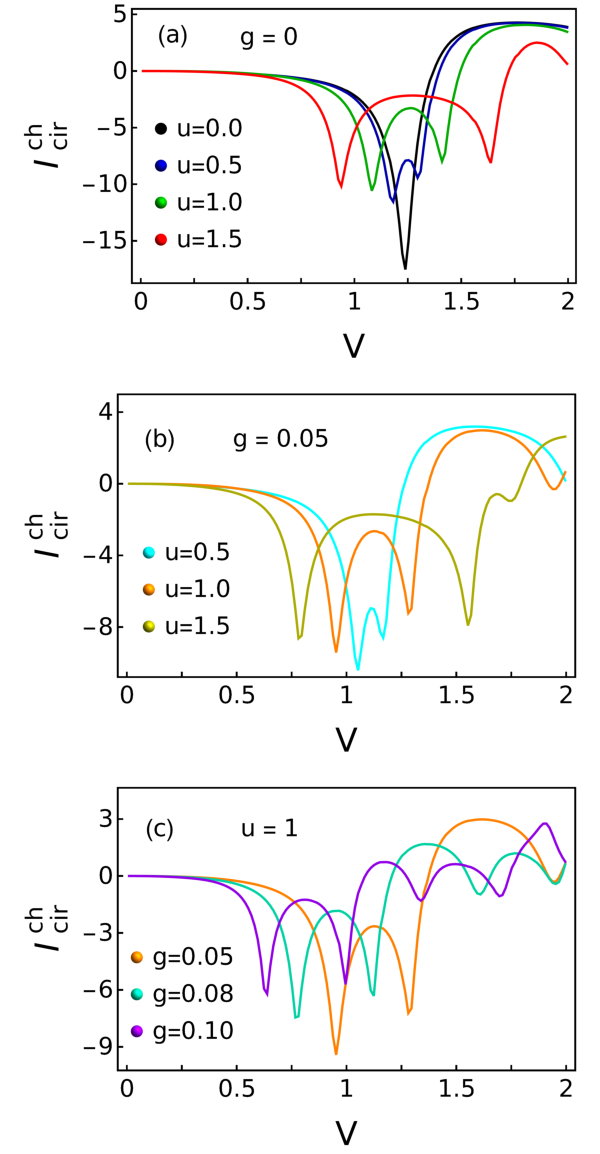}}
\caption{(Color online). Variation of the charge circular current $I^{ch}_{cir}$ with voltage $V$ for (a) increasing $u$ with $g=0$, 
(b) increasing $u$ with a constant non-zero $g$, and (c) constant $u$ with increasing $g$.}
\label{fig4}
\end{figure}
$u^{eff}<u$. The up and down spin spectra nevertheless remain different for the reasons discussed above.

Next, we plot the up and down spin circular currents as a function of the bias voltage in Fig.~\ref{fig3}. The Fermi energy is chosen 
such that the circular current density profile is approximately symmetric about the Fermi energy. The placement of the Fermi energy plays 
a crucial role in determining the current response of such systems. In Figs.~\ref{fig3}(a) and (d), the black curves correspond to the 
circular current of a perfect non-interacting ring. They exhibit a sharp dip over a narrow voltage range within the Fermi window, while 
the circular current remains negligibly small at other bias voltages. The asymmetry between the contributions from the peaks and dips 
within the bias window results in a non-zero circular current, indicating that, among the $+k$ and $-k$ states associated with a 
particular degenerate eigenvalue, one state contributes more significantly than the other.
The curves other than the black ones in Figs.~\ref{fig3}(a) and (d) represent the circular current in the presence of the e-e interaction. 
The Hubbard interaction causes an overall shift of the circular current density profile. For different values of $u$, the current density
profile is distributed over different energy windows, and the Fermi energy consequently shifts accordingly. As a result, a larger number 
of asymmetric peaks and dips can fall within the bias window at a given voltage, leading to a non-zero circular current over a wider 
voltage range compared with the non-interacting case. Another notable feature is that the threshold voltage decreases with increasing $u$.
\begin{figure}[ht]
\centering \resizebox*{7cm}{11.5cm}{\includegraphics{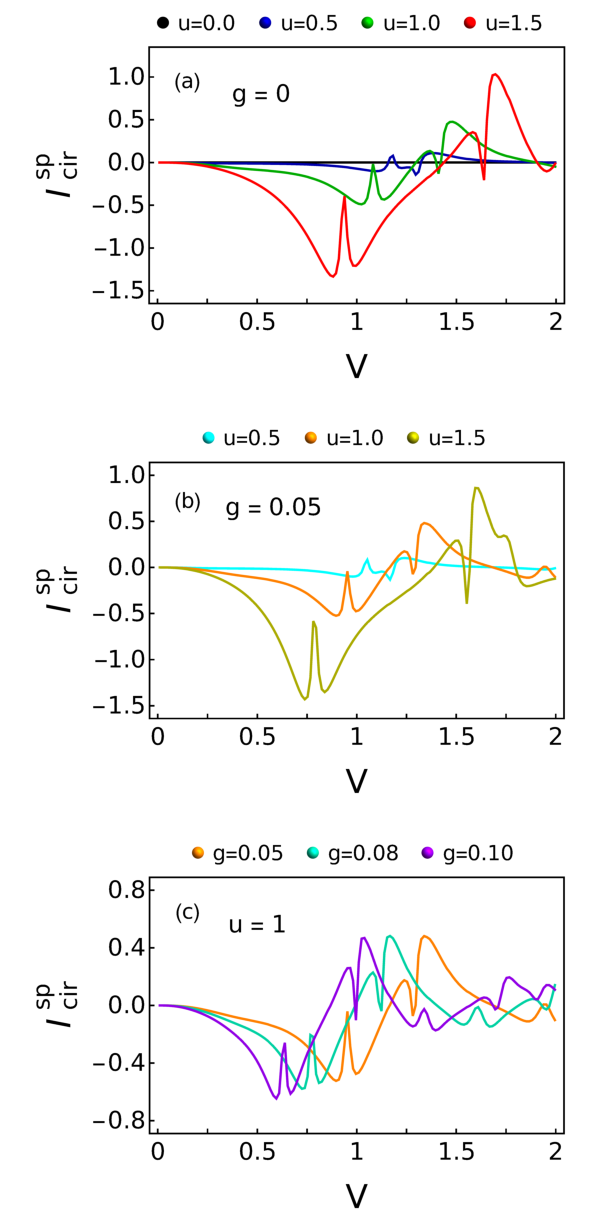}}
\caption{(Color online). Circular spin current $I^{sp}_{cir}$ as a function of bias voltage $V$ for three different combinations of 
$u$ and $g$, corresponding to the cases considered in Fig.~\ref{fig4}.}
\label{fig5}
\end{figure}
A similar trend is observed in Figs.~\ref{fig3}(b) and (e), where the circular current is calculated for increasing values of $u$ in the
presence of a fixed non-zero e-ph interaction. In Figs.~\ref{fig3}(c) and (f), increasing the e-ph interaction significantly reduces the
magnitude of the circular current. This reduction can be attributed to the effective narrowing of the bandwidth with increasing e-ph 
coupling. Consequently, a larger number of peaks and dips are incorporated within the Fermi window, resulting in two notable effects. 
First, a reduction in the current magnitude due to the enhanced mutual cancellation of the contributions from different energy ranges, 
and second, the appearance of a non-zero circular current over several bias voltages. The threshold voltage again decreases with increasing 
$g$. A closer examination of Fig.~\ref{fig3} further shows that our junction configuration gives rise to a circular charge current 
accompanied by a circular spin current. For a given set of $u$ and $g$, the up and down spin circular currents exhibit nearly identical 
voltage dependences, with opposite signs over the corresponding bias-voltage range.

The charge circular current is obtained by summing the up and down spin circular currents. As expected, the magnitude of the charge 
circular current is approximately twice that of the individual spin-resolved circular currents over most of the bias-voltage range, 
as shown in Fig.~\ref{fig4}. The trends observed in Figs.~\ref{fig4}(a)-(c) can be understood from the same considerations discussed 
for Fig.~\ref{fig3}. The sign of the circular current in Figs.~\ref{fig3} and \ref{fig4} depends on the applied bias. A positive 
(negative) sign denotes a clockwise (anti-clockwise) circulating current.

Figure~\ref{fig5} illustrates the circular spin current as a function of bias voltage. 
\begin{figure}[ht]
\centering \resizebox*{7cm}{8.8cm}{\includegraphics{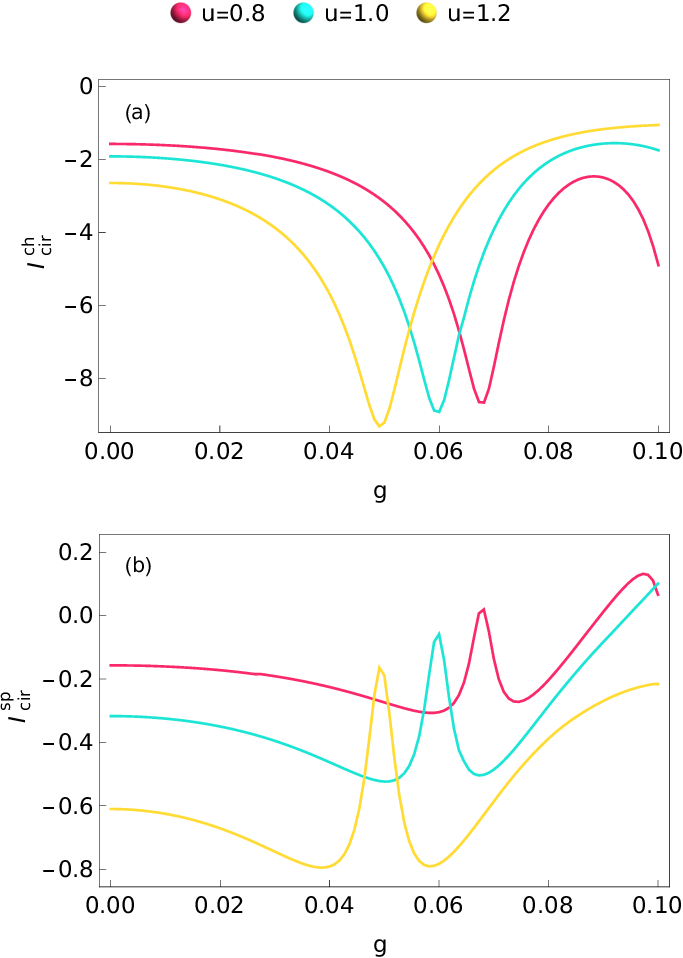}}
\caption{(Color online). (a) Circular charge current ($I^{ch}_{cir}$) and (b) circular spin current ($I{^{sp}_{cir}}$) as a function of 
$g$ for $V=0.9\,$V and three different values of $u$.}
\label{fig6}
\end{figure}
The circular spin current is obtained from the 
difference between the up and down spin circular currents. Here, a positive (negative) sign indicates that the circular current is 
predominantly carried by up spin (down spin) electrons. Figure~\ref{fig3} shows that, at certain bias voltages, the up and down spin 
circular currents exhibit a mismatch, which can also be observed in the corresponding current-density profiles in Fig.~\ref{fig2}. 
These bias voltages therefore give rise to a non-zero circular spin current.
In Fig.~\ref{fig5}(a), the circular spin current vanishes for the non-interacting case (black curve), since the up and down spin 
Hamiltonians are identical, with all site energies set to zero. In contrast, all other curves corresponding to finite values of $u$ 
exhibit a non-zero circular spin current. These curves show an overall enhancement of the circular spin current with increasing $u$. 
As $u$ increases while $t$ is kept fixed, the separation between the up and down spin DOS increases due to the enhanced spin-dependent
scattering. Consequently, the difference between the up and down spin circular currents increases, leading to an enhancement of 
$I^{sp}_{cir}$ with increasing $u$. A similar enhancement of the circular spin current with increasing $u$ is observed in 
Fig.~\ref{fig5}(b) even in the presence of a finite e-ph interaction strength. 
\begin{figure}[ht]
\centering \resizebox*{7cm}{11.5cm}{\includegraphics{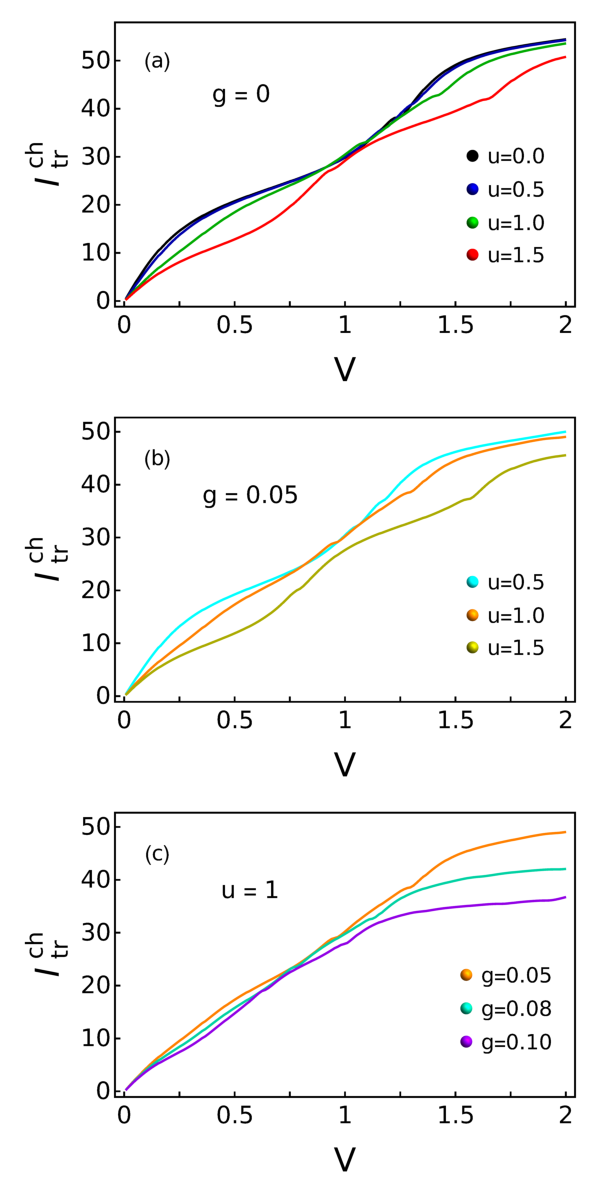}}
\caption{(Color online). Charge transport current ($I^{ch}_{tr}$) as a function of bias voltage $V$ for some typical values of $u$ 
and $g$, as specified within each panel.}
\label{fig7}
\end{figure}
In Fig.~\ref{fig5}(c), an overall enhancement of 
$I^{sp}_{cir}$ is observed with increasing $g$. With increasing $g$, both the effective Hubbard interaction $u^{eff}$ and the 
effective hopping strength $t^{eff}$ decrease. However, for small values of $g$, the ratio $u^{eff}/t^{eff}$ increases relative to 
its value at $g=0$ over a certain range, which effectively enhances the separation between the up and down spin current contributions 
within the ring. This accounts for the observed enhancement of the circular spin current with increasing $g$. The oscillatory behavior 
of the curves in Figs.~\ref{fig5}(a)-(c) indicates that the dominant contribution to the circular current can alternate between up 
and down spin electrons depending on the applied bias.

To further explore the effect of the e-ph coupling on the circular currents, we plot the circular charge current ($I^{ch}_{cir}$) 
and circular spin current ($I^{sp}_{cir}$) as a function of $g$ for a fixed bias voltage and three different values of $u$ in
Figs.~\ref{fig6}(a) and (b), respectively. The results demonstrate that the e-ph coupling strength $g$ provides an effective tuning 
parameter for both the circular charge and spin currents. In particular, both currents can be enhanced by appropriately tuning the 
value of $g$.

Let us now focus our attention on the transport currents in our ring junction setup and investigate their behavior one by one for 
different values of the interaction parameters. 

Figure~\ref{fig7} displays the charge transport current, $I^{ch}_{tr}$, as a function of the bias voltage $V$. Figures~\ref{fig7}(a) 
and (b) show the charge transport current for different values of the Hubbard interaction strength $u$. 
\begin{figure}[ht]
\centering \resizebox*{7cm}{11.5cm}{\includegraphics{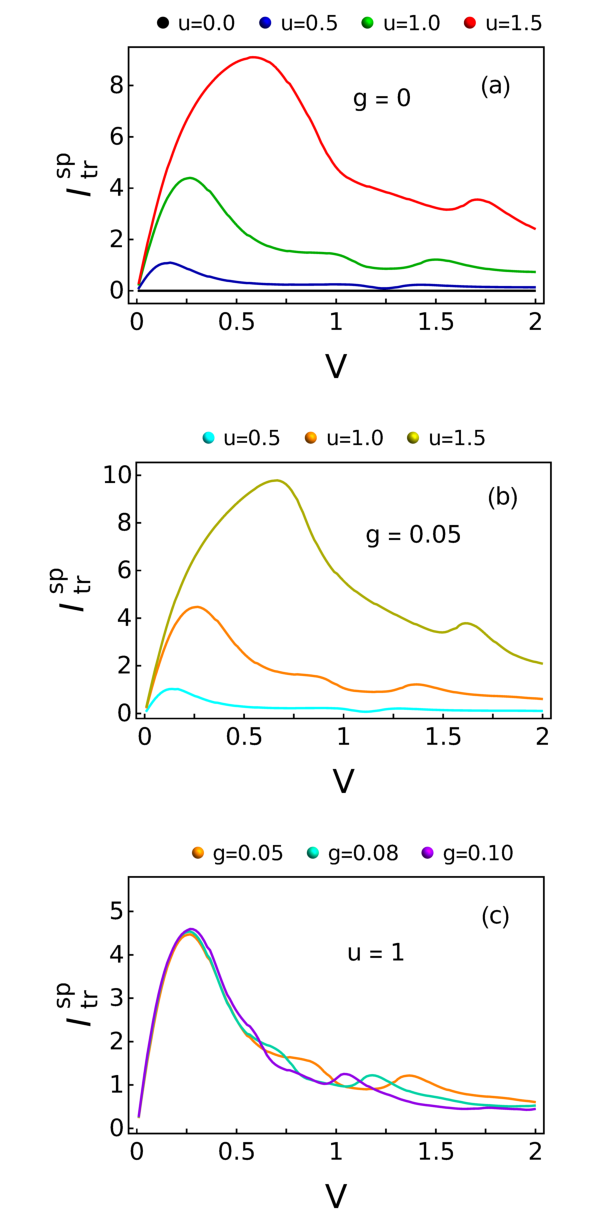}}
\caption{(Color online). Spin transport current ($I{^{sp}_{tr}}$) as a function of bias voltage $V$ for different combinations 
of $u$ and $g$, as specified within each panel.}
\label{fig8}
\end{figure}
In both panels, the current 
gradually decreases with increasing $u$. This behavior can be understood from the corresponding electronic structure. At half-filling, 
the Hubbard interaction favors an antiferromagnetic ordering, and the resulting spin-dependent Hamiltonians resemble those of a 
bipartite lattice, leading to the opening of a gap around the band center. At absolute zero temperature, the Fermi energy is positioned 
at the highest occupied level of the lower band. Since all states below the Fermi energy are occupied at half-filling, the presence 
of the gap suppresses electron transmission within the low-bias transport window. As the Hubbard interaction strength increases, the 
gap becomes wider, thereby reducing the number of available conducting states within the bias window and consequently suppressing the 
charge transport current. This behavior is consistent with the tendency toward an insulating state in the regime of strong e-e correlations.
In Fig.~\ref{fig7}(c), the e-ph coupling strength $g$ is varied while keeping the e-e interaction strength $u$ fixed. Increasing $g$ leads 
to a reduction in the effective hopping strength and, consequently, a narrowing of the electronic bandwidth. 
\begin{figure}[ht]
\centering \resizebox*{7.5cm}{5cm}{\includegraphics{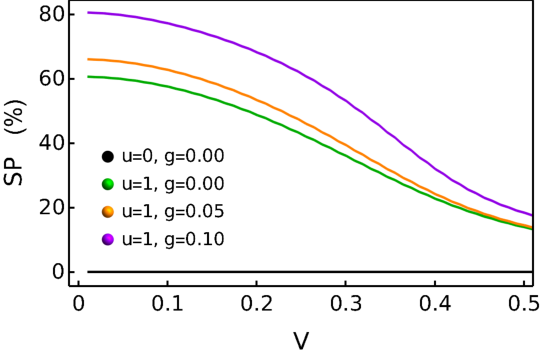}}
\caption{(Color online). Spin polarization (SP) coefficient as a function of bias voltage $V$ for four different sets of $u$ and $g$, 
as described in the figure using four distinct colors.}
\label{fig9}
\end{figure}
As a result, the transmission
resonances become confined to a narrower energy range, which reduces their overall contribution to the transport current. In addition, 
the effective coupling between the ring and the electrodes is reduced in the presence of the e-ph interaction, further suppressing electron
transfer across the junction. These effects provide a microscopic basis for the observed decrease in the low-bias current with increasing 
$g$. 
\begin{figure}[ht]
\centering \resizebox*{7.5cm}{4.5cm}{\includegraphics{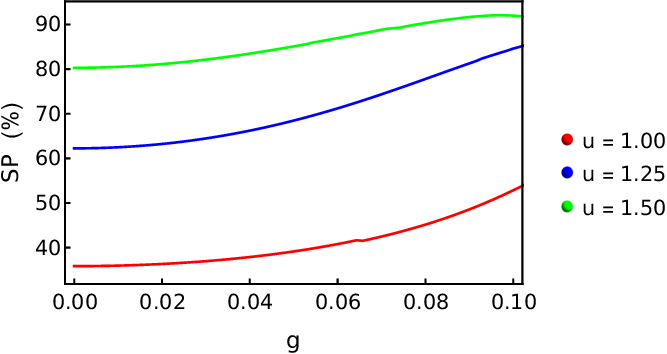}}
\caption{(Color online). Spin polarization (SP) coefficient as a function of e-ph coupling strength $g$ for three different values 
of $u$ at $V=0.3\,$V.} 
\label{fig10}
\end{figure}
The suppression of low-bias transport can also be viewed qualitatively in terms of the Franck-Condon blockade. In the presence of 
e-ph coupling, electron tunneling is accompanied by a change in the vibrational state of the conductor. Consequently, the tunneling 
probability depends not only on the overlap of the electronic wave functions but also on the overlap between the corresponding vibrational
states. The displacement of the vibrational potential induced by the e-ph coupling reduces the overlap between the relevant low-lying
vibrational states, thereby suppressing low-bias tunneling. Thus, the e-ph interaction can lead to phonon-assisted suppression of electron
transport, particularly in the low-bias regime.

Similarly, the spin transport current, $I^{sp}_{tr}$, as a function of the bias voltage $V$ is illustrated in Fig.~\ref{fig8}. The spin
transport current is defined as the difference between the up and down spin transport currents and therefore depends on the difference 
between their corresponding transmission spectra. 
\begin{figure}[ht]
\centering \resizebox*{8cm}{7.5cm}{\includegraphics{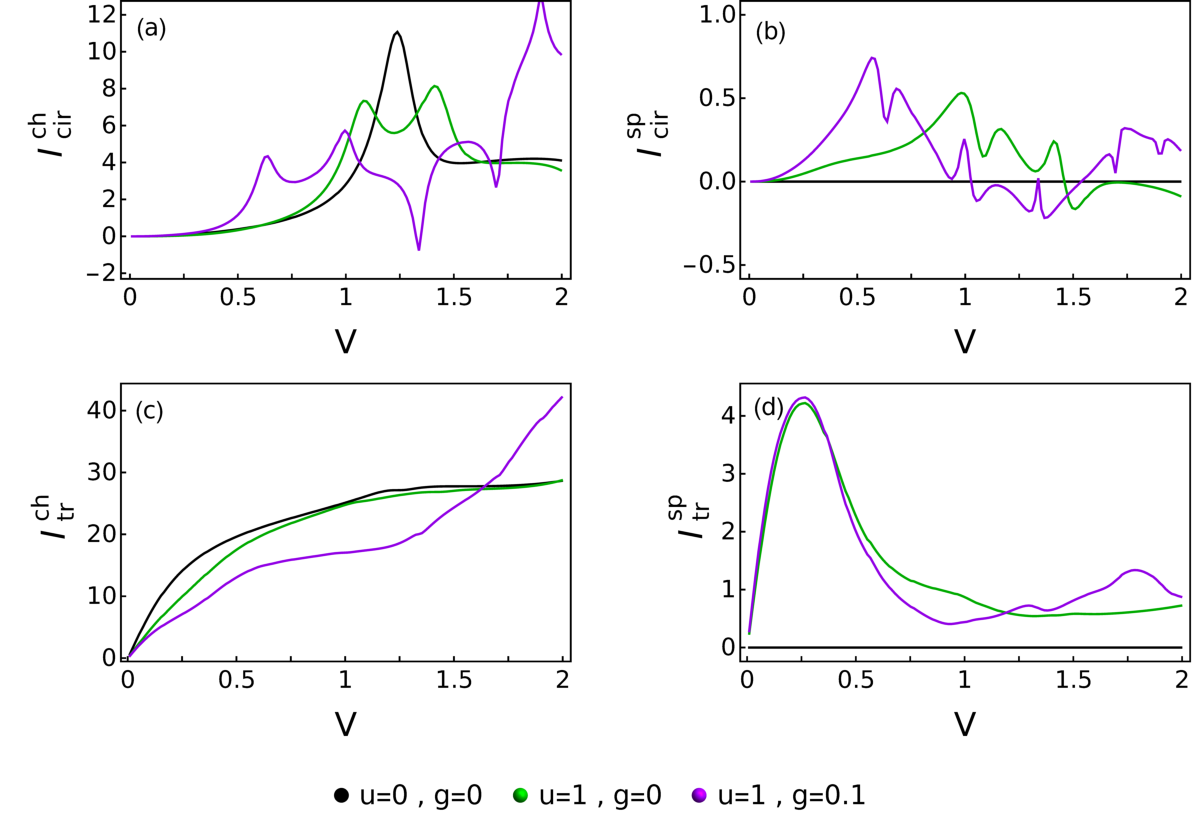}}
\caption{(Color online). Charge and spin currents as a function of bias voltage for a different junction configuration, with the source 
and drain connected to sites $1$ and $15$ of the ring, respectively. (a) Circular charge current ($I^{ch}_{cir}$), (b) circular spin 
current ($I^{sp}_{cir}$), (c) transport charge current ($I^{ch}_{tr}$), and (d) transport spin current ($I^{sp}_{tr}$).}
\label{fig11}
\end{figure}
In Fig.~\ref{fig8}(a), a non-zero spin transport current appears only when the e-e 
interaction is present. In the presence of the Hubbard interaction, the coupling to the leads breaks the sublattice symmetry between the 
up and down spin Hamiltonians, resulting in different spin-dependent transmission characteristics and, consequently, a finite spin 
transport current. The spin transport current increases with increasing Hubbard interaction strength, as observed in Figs.~\ref{fig8}(a) 
and (b) for zero and finite values of $g$, respectively. With increasing $u$, the antiferromagnetic ordering becomes more pronounced, 
leading to an enhancement of the local magnetic moments and, consequently, stronger spin-dependent scattering of the itinerant electrons. 
This enhances the difference between the up and down spin transport currents and results in an increase in the magnitude of $I^{sp}_{tr}$. 
A positive (negative) value of the spin transport current indicates that the transport current is predominantly carried by up spin 
(down spin) electrons. In Fig.~\ref{fig8}(c), the spin transport current shows an overall enhancement with increasing e-ph coupling 
strength $g$ for a fixed value of $u$. This behavior can be understood from the dependence of the effective interaction parameters on $g$. 

For fixed $u$ and $t$, the degree of spin-dependent response is governed, in part, by the relative strength of the e-e interaction 
with respect to the hopping amplitude. In the presence of e-ph coupling, both the effective Hubbard interaction $u^{eff}$ and the effective
hopping strength $t^{eff}$ are reduced. However, $t^{eff}$ decreases exponentially with increasing $g$, whereas $u^{eff}$ decreases linearly.
Consequently, over a certain range of $g$, the ratio $u^{eff}/t^{eff}$ increases relative to its value at $g=0$. This effectively enhances 
the spin-dependent scattering and increases the difference between the up and down spin transport currents, leading to the observed 
enhancement of $I^{sp}_{tr}$ with increasing $g$.
\begin{figure}[ht]
\centering \resizebox*{8cm}{7.5cm}{\includegraphics{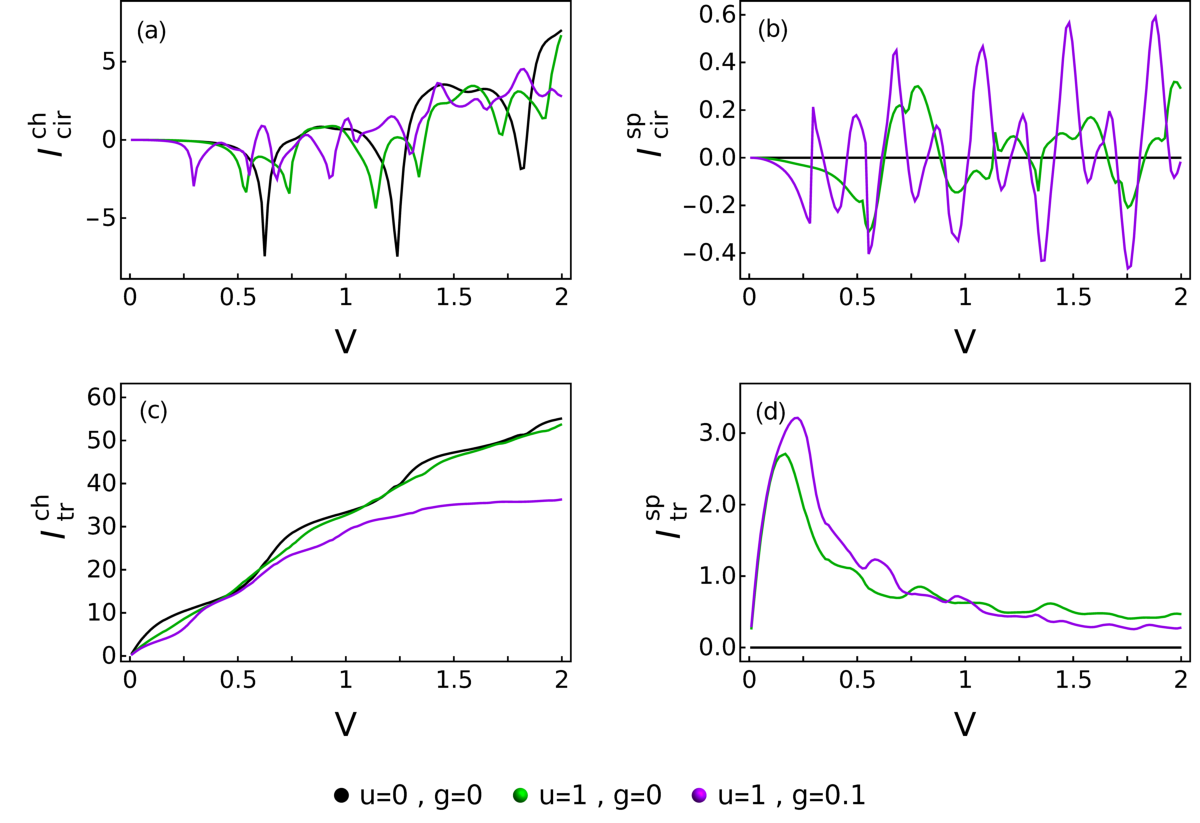}}
\caption{(Color online). System-size dependence of the charge and spin currents as functions of bias voltage for a $40$-site conducting 
ring, with the source and drain connected to sites $1$ and $19$ of the ring, respectively.}
\label{fig12}
\end{figure}
The above features can be understood more clearly from the spin polarization (SP) coefficient as a function of bias voltage, shown in
Fig.~\ref{fig9}. In the non-interacting case ($u=0$ and $g=0$), no spin polarization occurs because the up and down spin transport channels 
are identical. For a fixed value of $u$, the spin polarization increases with increasing $g$. This behavior is consistent with the trend
observed for the spin transport current in Fig.~\ref{fig8}. The introduction of e-ph coupling modifies the effective electronic parameters 
such that, over the relevant range of $g$, the ratio $u^{eff}/t^{eff}$ becomes larger than $u/t$. This enhances the relative
strength of the effective e-e interaction compared with the effective hopping and consequently increases the spin-dependent separation of 
the transport characteristics in the Hubbard-like system.
It is also noteworthy that the spin transport current and spin polarization nearly vanish at higher bias voltages in both Figs.~\ref{fig9} 
and \ref{fig10}. As the bias voltage increases, the transport windows for the up and down spin channels progressively overlap, allowing
comparable contributions from both spin species. Their contributions to the spin transport current therefore tend to compensate each other,
resulting in a strongly reduced net spin current and, consequently, a nearly vanishing spin polarization. For further clarification,
Fig.~\ref{fig10} shows the spin polarization as a function of $g$ at a fixed bias voltage. For each value of $u$ considered, the spin
polarization increases monotonically with increasing $g$. This behavior is consistent with the increase in the effective 
interaction-to-hopping ratio discussed above. In Figs.~\ref{fig8}, \ref{fig9}, and \ref{fig10}, the spin transport current and spin 
polarization remain positive for the parameter ranges considered. Thus, for the particular junction configuration studied here, the 
transport current is predominantly carried by up-spin electrons.

We know that the junction configuration can play an important role in determining the transport characteristics of open quantum systems.
Figure~\ref{fig11} presents the $I$-$V$ characteristics of all four currents, $I_{cir}^{ch}$, $I_{cir}^{sp}$, $I_{tr}^{ch}$, and
$I_{tr}^{sp}$, for a different junction configuration, where the source and drain are connected to the $1$st and $15$th sites of the 
ring, respectively. The qualitative features of the current responses remain similar to those obtained for the junction configuration 
considered above. In particular, the dependence of the charge and spin currents on the bias voltage and the interaction parameters 
remains qualitatively unchanged. This indicates that the principal features of the current responses are not specific to the particular 
junction configuration considered previously.

Finally, in Fig.~\ref{fig12}, we check the system-size dependence of the obtained results by considering a $40$-site ring, with the 
source and drain connected to sites $1$ and $19$, respectively. All other parameters are kept the same as in the preceding calculations. 
The charge and spin currents exhibit trends with the e-e interaction strength $u$ and the e-ph coupling strength $g$ that are qualitatively
consistent with the results obtained for the smaller ring. Thus, the principal interaction-dependent features of both the circular and 
transport currents persist upon increasing the system size. 

\section{Closing Remarks}

To conclude, we have investigated bias-driven circular charge and spin currents in a ring nanojunction in the presence of e-e and e-ph
interactions within a TB framework based on the NEGF formalism. The interacting system is mapped onto an effective electronic model through 
the Lang-Firsov transformation, which is subsequently treated within the Hartree-Fock mean-field scheme. Using this framework, we have
systematically investigated both circular and transport charge and spin currents over a broad range of interaction strengths and bias 
voltages, and have also explored the sensitivity of the results to the junction configuration and system size.

Our numerical results demonstrate that bias-driven circular currents can be significantly enhanced over a relatively broad range of bias
voltages. A finite spin current can also be generated within the ring, with the dominant contribution arising from either up spin or down
spin electrons depending on the applied bias and interaction strengths. The e-e and e-ph interactions play important and distinct roles 
in determining the charge and spin responses. While the charge transport current is suppressed with increasing interaction strengths, 
owing to effects such as the interaction-induced opening of a gap around the band center and the suppression of low-bias transport 
associated with e-ph coupling, the spin-dependent response becomes more pronounced. In particular, the spin polarization can be enhanced 
by increasing the e-e interaction strength, the e-ph coupling strength, or both, over the parameter ranges considered here. The principal
features of the circular and transport currents also persist for different junction configurations and for a larger ring, indicating that 
the observed behavior is not restricted to a particular contact geometry or system size.

Overall, our theoretical findings based on a model ring nanojunction provide a framework for realizing and controlling selective spin 
transport through the interplay of e-e and e-ph interactions under an applied voltage bias. The present study may offer useful insights 
into the interplay between interaction effects, junction geometry, and charge and spin transport in nanojunctions possessing single 
and multiple loops sub-structures.

\section*{ACKNOWLEDGMENTS}

The authors sincerely thank Prof. S. Sil for valuable discussions.

\section*{DATA AVAILABILITY STATEMENT}

The data supporting the findings of this study are included in the manuscript.


\section*{DECLARATION}

{\bf Conflict of interest} The authors declare no conflict of interest.

\appendix
\section{Lang-Firsov transformation}
\label{ap1}

Here, we explicitly derive the effective electronic ring Hamiltonian by eliminating the e-ph coupling term. Let us begin with the full
ring Hamiltonian in the presence of both the e-e and e-ph interactions
\begin{eqnarray}
H_R&=&\sum_{i,\sigma} \epsilon_{i\sigma} n_{i\sigma}  + t\sum_{<i,j>,\sigma} \left[c{_{i\sigma}^{\dagger}} c_{j\sigma} + 
c{_{j\sigma}^{\dagger}} c_{i\sigma}\right]\nonumber\\
& + & u\sum_{i} n_{i\uparrow} n_{i\downarrow} + \hbar \omega_{0} \sum_i b{{_i}{^\dagger}}b_i \nonumber\\ 
& + & g\sum_{i,\sigma} (b{{_i}{^\dagger}}+ b_i) n_{i\sigma}
\label{eqn23}
\end{eqnarray}
where $n_{i\sigma}$ ($=c{_{i\sigma}^{\dagger}}c{_{i\sigma}}$) is the number operator. We choose the transformation generator as
\begin{eqnarray}
U=\left(\frac{g}{\hbar \omega_{0}}\right) \sum_{i,\sigma} \left(b{_{i}^{\dagger}}-b_{i}\right) n_{i\sigma}.
\end{eqnarray}
Using this generator, the ring Hamiltonian can be transformed as
\begin{eqnarray}
\widetilde{H}_R & = & e^U\,\,H_R\,\,e^{-U}\nonumber\\
& = & A\,H_{R}\,A^{-1} \nonumber ~~~~ (\mbox{where}~ A=e^{U}) \nonumber\\
& = & \sum_{i,\sigma} \epsilon_{i\sigma}A\left(c{^{\dagger}_{i\sigma}}c_{i\sigma}\right)A^{-1}\nonumber\\
& + & t \sum_{<i,j>,\sigma} A\left[c{_{i\sigma}^{\dagger}}c_{j\sigma}+h.c\right]A^{-1}\nonumber\\
& + & u\sum_{i} A\left(c{_{i\uparrow}^{\dagger}}c_{i\uparrow}c{_{i\downarrow}^{\dagger}}c_{i\downarrow}\right)A^{-1}\nonumber\\
& + & \hbar\omega_{0}\,\sum_{i} A\left(b{_{i}^{\dagger}}b_{i}\right)A^{-1} \nonumber\\ 
& + & g\sum_{i,\sigma} A\left(b{_i^{\dagger}}+b_i\right)c{_{i\sigma}^{\dagger}}c_{i\sigma}A^{-1}.
\end{eqnarray}
More elaborately, we can write
\begin{eqnarray}
\widetilde{H}_R & =& \sum_{i,\sigma} \epsilon_{i\sigma}Ac{^{\dagger}_{i\sigma}}A^{-1}Ac_{i\sigma}A^{-1}\nonumber\\
& + & t \sum_{<i,j>,\sigma} \left[Ac{_{i\sigma}^{\dagger}}A^{-1}Ac_{j\sigma}A^{-1}+h.c\right]\nonumber\\
& + & u\sum_{i} Ac{_{i\uparrow}^{\dagger}}A^{-1}Ac_{i\uparrow}A^{-1}Ac{_{i\downarrow}^{\dagger}}A^{-1}Ac_{i\downarrow}A^{-1}\nonumber\\
& + & \hbar\omega_{0}\,\sum_{i} Ab{_{i}^{\dagger}}A^{-1}Ab_{i}A^{-1} \nonumber\\ 
& + & g\sum_{i,\sigma} A\left(b{_i^{\dagger}}+b_i\right)A^{-1}Ac{_{i\sigma}^{\dagger}}A^{-1}Ac_{i\sigma}A^{-1}\nonumber\\
& = & \sum_{i,\sigma} \epsilon_{i\sigma}\left(\tilde{c}{^{\dagger}_{i\sigma}}\tilde{c}_{i\sigma}\right)+t \sum_{<i,j>,\sigma}\left[\tilde{c}{_{i\sigma}^{\dagger}}\tilde{c}_{j\sigma}+h.c\right]\nonumber\\
& + & u\sum_{i}\left(\tilde{c}{_{i\uparrow}^{\dagger}}\tilde{c}_{i\uparrow}\tilde{c}{_{i\downarrow}^{\dagger}}\tilde{c}_{i\downarrow}\right)
+ \hbar\omega_{0}\,\sum_{i}\left(\tilde{b}{_{i}^{\dagger}}\tilde{b}_{i}\right) \nonumber\\
& + & g\sum_{i,\sigma} \left(\tilde{b}{_i^{\dagger}}+\tilde{b}_i\right) \tilde{c}{_{i\sigma}^{\dagger}}\tilde{c}_{i\sigma}
\label{eqn24}
\end{eqnarray}
where, 
\begin{eqnarray}
\tilde{c}{_{i\sigma}^{\dagger}} = Ac{_{i\sigma}^{\dagger}}A^{-1},~~\tilde{c}_{i\sigma} & = & Ac_{i\sigma}A^{-1} \nonumber \\
\tilde{b}{_{i}^{\dagger}} = Ab{_i^{\dagger}}A^{-1},~~~~~\tilde{b}_{i}& = & Ab{_i}A^{-1}. 
\label{ap1eq4}
\end{eqnarray}
We now derive the four operators given in Eq.~\ref{ap1eq4}, one by one, using the Baker-Campbell-Hausdroff (BCH) relation as follows.
For any general operator `B', the BCH relation is 
\begin{eqnarray}
e^ABe^{-A} & = & \sum_{n=0}^{\infty} \frac{1}{n!}\left[A,B\right]_n \nonumber \\
& = & B+\left[A,B\right]+\frac{1}{2}\left[A,\left[A,B\right]\right] \nonumber\\
& + & \frac{1}{6} \left[A,\left[A,\left[A,B\right]\right]\right]+\dots
\end{eqnarray}
Hence,
\begin{eqnarray}
\tilde{c}{_{i\sigma}^{\dagger}} & = & A c{_{i\sigma}^{\dagger}}A^{-1} \nonumber \\
& = & e^U\, c{_{i\sigma}^{\dagger}}\,e^{-U} \nonumber \\
& = & c{_{i\sigma}^{\dagger}}+\left[U,c{_{i\sigma}^{\dagger}}\right]+\frac{1}{2}\left[U,\left[U,c{_{i\sigma}^{\dagger}}\right]\right]+\dots 
\label{ap1eq7}
\end{eqnarray}
The different commutation brackets of Eq.~\ref{ap1eq7} can be calculated as follows.
\begin{eqnarray}
\left[U,c{_{i\sigma}^{\dagger}}\right] & = & \left(\frac{g}{\hbar\omega_0}\right)\sum_{i,\sigma}\left(b{_{i}^{\dagger}}-b_{i}\right)
\underbrace{\left[n_{i\sigma}\,,\,c{_{i\sigma}^{\dagger}}\right]}_{\eta} \nonumber\\
& = &\left(\frac{g}{\hbar\omega_0}\right)\left(b{_{i}^{\dagger}}-b_{i}\right)c{_{i\sigma}^{\dagger}}.
\label{eqn25}
\end{eqnarray}
\begin{eqnarray}
\left[U,\left[U,c{_{i\sigma}^{\dagger}}\right]\right] & = & \left(\frac{g}{\hbar\omega_0}\right)^2\left(b{_{i}^{\dagger}}-b_{i}\right)^2
c{_{i\sigma}^{\dagger}}. 
\label{eqn26}
\end{eqnarray}
$\square$ Calculation of $\eta$:
\begin{eqnarray}
\left[n_{i\sigma}\,,c{^{\dagger}_{i\sigma}}\right] & = &n_{i\sigma} c{_{i\sigma}^{\dagger}}-c{_{i\sigma}^{\dagger}}n_{i\sigma}\nonumber\\
&= & c{_{i\sigma}^{\dagger}}c_{i\sigma}c{_{i\sigma}^{\dagger}}-c{_{i\sigma}^{\dagger}}c{_{i\sigma}^{\dagger}}c_{i\sigma}\nonumber\\
&= &c{_{i\sigma}^{\dagger}}\left(1-c{_{i\sigma}^{\dagger}}c_{i\sigma}\right)-c{_{i\sigma}^{\dagger}}c{_{i\sigma}^{\dagger}}c_{i\sigma}\nonumber\\
& = & c{_{i\sigma}^{\dagger}}-2c{_{i\sigma}^{\dagger}}c{_{i\sigma}^{\dagger}}c_{i\sigma}\nonumber\\
& = &c{_{i\sigma}^{\dagger}}.
\end{eqnarray}
Using Eq.~\ref{eqn25} and Eq.~\ref{eqn26} we can write,
\begin{eqnarray}
\tilde{c}{_{i\sigma}^{\dagger}} & = & \exp\left[\left(\frac{g}{\hbar\omega_0}\right)\left(b{_{i}^{\dagger}}-b_{i}\right)\right]
c{_{i\sigma}^{\dagger}},
\end{eqnarray}
which simplifies to
\begin{eqnarray}
\tilde{c}{_{i\sigma}^{\dagger}} & = & c{_{i\sigma}^{\dagger}}\exp\left[\left(\frac{g}{\hbar\omega_0}\right)\left(b{_{i}^{\dagger}}-b_{i}\right)\right].
\label{eqn27}
\end{eqnarray}
Simiarly,
\begin{eqnarray}
\tilde{c}_{i\sigma} & = & c_{i\sigma}\exp\left[-\left(\frac{g}{\hbar\omega_0}\right)\left(b{_{i}^{\dagger}}-b_{i}\right)\right].
\label{eqn28}
\end{eqnarray}
In the same footing we evaluate the phononic operators as follows.
\begin{eqnarray}
\tilde{b}_i^{\dagger} & = & e^U\,b{_i^{\dagger}}\,e^{-U}\nonumber\\
& = & b{_i^{\dagger}}+\left[U,b{_i^{\dagger}}\right]+\frac{1}{2}\left[U,\left[U,b{_i^{\dagger}}\right]\right]+\dots
\label{ap1eq14} 
\end{eqnarray}
Now, we compute the commutation brackets of Eq.~\ref{ap1eq14}.
\begin{eqnarray}
\left[U\,,{b}{_i^{\dagger}}\right] & = & \left(\frac{g}{\hbar\omega_0}\right)\sum_{i,\sigma} n_{i\sigma}\, \underbrace{\left[\left(b{_i^{\dagger}}-b_i\right)\, ,\,b{_i^{\dagger}}\right]}_{\zeta} \nonumber \\
& = & -\left(\frac{g}{\hbar\omega_0}\right)\sum_{\sigma} n_{i\sigma}.
\label{eqn29}
\end{eqnarray}
\begin{eqnarray}
\left[U\,,\,\left[U\,,\,b{_i^{\dagger}}\right]\right]=0.
\label{eqn30}
\end{eqnarray}
$\square$ Calculation of $\zeta$:
\begin{eqnarray}
\left[\left(b{_i^{\dagger}}-b_i\right)\, ,\,b{_i^{\dagger}}\right]=\left[b{_i^{\dagger}}\,,b{_i^{\dagger}}\right]-\left[b_i\,,b{_i^{\dagger}}\right]=-1.
\end{eqnarray}
Using Eq.~\ref{eqn29} and Eq.~\ref{eqn30}, we get 
\begin{eqnarray}
\tilde{b}{_i^{\dagger}} & = & b{_i^{\dagger}}-\left(\frac{g}{\hbar\omega_0}\right)\sum_{\sigma} n_{i\sigma}.
\label{eqn31}
\end{eqnarray}
Similary,
\begin{eqnarray}
\tilde{b}_i & = & b_i -\left(\frac{g}{\hbar\omega_0}\right)\sum_{\sigma} n_{i\sigma}.
\label{eqn32}
\end{eqnarray}
Substituting the compact forms of the operators derived in Eq.~\ref{eqn27}, Eq.~\ref{eqn28}, Eq.~\ref{eqn31}, and Eq.~\ref{eqn32}, we 
now evaluate different operator terms of the ring Hamiltonian. The $1$st term of $\widetilde{H}_R$ (see, Eq.~\ref{eqn24}) becomes   
\begin{eqnarray}
\tilde{c}{_{i\sigma}^{\dagger}}\tilde{c}_{i\sigma} & = & c{_{i\sigma}^{\dagger}}\exp\left[\left(\frac{g}{\hbar\omega_0}\right)\left(b{_i^{\dagger}}-b_i\right)\right]\nonumber\\
& &c_{i\sigma}\exp\left[-\left(\frac{g}{\hbar\omega_0}\right)\left(b{_i^{\dagger}}-b_i\right)\right]\nonumber\\
& =  & c{_{i\sigma}^{\dagger}}c_{i\sigma}\exp\left[\left(\frac{g}{\hbar\omega_0}\right)\left(b{_i^{\dagger}}-
b_i\right)\right]\nonumber\\
&   &\exp\left[-\left(\frac{g}{\hbar\omega_0}\right)\left(b{_i^{\dagger}}-b_i\right)\right]\nonumber\\
& = &  c{_{i\sigma}^{\dagger}}c_{i\sigma} \underbrace{\exp\left[\left(\frac{g}{\hbar\omega_0}\right)\left(b{_i^{\dagger}}-b_i-
b{_i^{\dagger}}-b_i\right)\right]}_\beta \nonumber\\
& = & c{_{i\sigma}^{\dagger}}c_{i\sigma}.
\label{eqn33}
\end{eqnarray}
$\square$ Calculation of $\beta$: We know for any two operators $A$ and $B$
\begin{eqnarray}
e^{A+B} & = & e^{-\frac{1}{2}\left[A\,,\,B\right]}\,e^A\,e^B. \\
\mbox{If}, A & = & \left(\frac{g}{\hbar\omega_0}\right)\left(b{_i^{\dagger}}-b_i\right),\nonumber\\
\mbox{and}~  B & = &-\left(\frac{g}{\hbar\omega_0}\right)\left(b{_i^{\dagger}}-b_i\right),\nonumber\\
\end{eqnarray}
then,
\begin{eqnarray}
\left[A\,,\,B\right] & = & \left(\frac{g}{\hbar\omega_0}\right)^2\left[\left(b{_i^{\dagger}}-b_i\right)\,,\,\left(b_i-b{_i^{\dagger}}\right)\right]\nonumber\\
& = & \left(\frac{g}{\hbar\omega_0}\right)^2\bigg(\left[b{_i^{\dagger}}\,,\,b_i\right]-\left[b_i\,,\,b_i\right]\nonumber\\
& - &\left[b{_i^{\dagger}}\,,\,b{_i^{\dagger}}\right]+\left[b_i\,,\,b{_i^{\dagger}}\right]\bigg)\nonumber\\
& = & 0. 
\end{eqnarray}
So, for these operator forms of $A$ and $B$,
\begin{equation}
e^{A+B}=e^A\,e^B.
\end{equation}
Now, evaluate the $2$nd term of $\widetilde{H}_R$ (see, Eq.~\ref{eqn24}).
\begin{eqnarray}
\tilde{c}_{j\sigma} & = & e^U\, c{_{j\sigma}}\,e^{-U}\nonumber\\
& = & c_{j\sigma}+\left[U,c{_{j\sigma}}\right]+\frac{1}{2}\left[U,\left[U,c{_{j\sigma}}\right]\right]+\dots
\label{ap1eq25}
\end{eqnarray}
The commutation bracket of Eq.~\ref{ap1eq25} is:
\begin{eqnarray}
\left[U,c{_{j\sigma}}\right]&=&\left(\frac{g}{\hbar\omega_0}\right)\sum_{i,\sigma} \left(b{_i^{\dagger}}-b_i\right)\left[c{_{i\sigma}^{\dagger}}
c_{i\sigma}\,,\,c{_{j\sigma}}\right]\nonumber\\
&=&\left(\frac{g}{\hbar\omega_0}\right)\sum_{i,\sigma} \left(b{_i^{\dagger}}-b_i\right) \bigg\{c{_{i\sigma}^{\dagger}}c_{i\sigma}c{_{j\sigma}}-c{_{j\sigma}}
c{_{i\sigma}^{\dagger}}c_{i\sigma}\bigg\}\nonumber\\
&=&\left(\frac{g}{\hbar\omega_0}\right)\sum_{i,\sigma} \left(b{_i^{\dagger}}-b_i\right) \bigg\{-c{_{i\sigma}^{\dagger}}
c{_{j\sigma}}c_{i\sigma}-c{_{j\sigma}}c{_{i\sigma}^{\dagger}}c_{i\sigma}\bigg\}\nonumber\\
&=&\left(\frac{g}{\hbar\omega_0}\right)\sum_{i,\sigma} \left(b{_i^{\dagger}}-b_i\right) \nonumber\\
&&\bigg\{-\left(\delta_{i,j}-c{_{j\sigma}}c{_{i\sigma}^{\dagger}}\right)c_{i\sigma}-c{_{j\sigma}}
c{_{i\sigma}^{\dagger}}c_{i\sigma}\bigg\}\nonumber\\
&= &\left(\frac{g}{\hbar\omega_0}\right) \sum_{i,\sigma} \left(b{_i^{\dagger}}-b_i\right)\bigg\{-\delta_{i,j}c_{i\sigma}\bigg\}\nonumber\\
&=&-\left(\frac{g}{\hbar\omega_0}\right) \left(b{_j^{\dagger}}-b_j\right)c_{j\sigma}.
\end{eqnarray}
Hence, the 2nd operator term of $\widetilde{H}_R$ becomes  
\begin{eqnarray}
\left[\tilde{c}{_{i\sigma}^{\dagger}}\tilde{c}_{j\sigma}+h.c\right] & = & c{_{i\sigma}^{\dagger}}c_{j\sigma}
\exp\left[\left(\frac{g}{\hbar\omega_0}\right)\left(b{_i^{\dagger}}-b_i\right)\right]\nonumber\\
&   &\exp\left[-\left(\frac{g}{\hbar\omega_0}\right)\left(b{_j^{\dagger}}-b_j\right)\right]\nonumber\\
&+ & h.c.
\label{eqn34}
\end{eqnarray}
The $3$rd term of Eq.~\ref{eqn24} simplifies to
\begin{eqnarray}
\tilde{c}{_{i\uparrow}^{\dagger}}\tilde{c}_{i\uparrow}\tilde{c}{_{i\downarrow}^{\dagger}}\tilde{c}_{i\downarrow} & = & c{_{i\uparrow}^{\dagger}}
c_{i\uparrow}c{_{i\downarrow}^{\dagger}}c_{i\downarrow}
\label{eqn35}
\end{eqnarray}
The $4$th term of of the ring Hamiltonian (Eq.~\ref{eqn24}) is obtained as follows.
\begin{eqnarray}
\hbar\omega_0\sum_i \tilde{b}{_i^{\dagger}}\tilde{b}_i & = & \hbar\omega_0\sum_i \left[b{_i^{\dagger}}-\left(\frac{g}{\hbar\omega_0}\right)\sum_{\sigma} n_{i\sigma}\right]\nonumber\\
&   & \left[b{_i}-\left(\frac{g}{\hbar\omega_0}\right)\sum_{\sigma} n_{i\sigma}\right]\nonumber\\
& = & \hbar\omega_0\sum_i b{_i^{\dagger}}b_i - \hbar\omega_0\left(\frac{g}{\hbar\omega_0}\right)\sum_{i,\sigma}
\left(b{_i^{\dagger}}+b_i\right)n_{i\sigma}\nonumber\\
& +  & \hbar\omega_0\left(\frac{g}{\hbar\omega_0}\right)^2\sum_{i,\sigma,\sigma^{\prime\prime}}
n_{i\sigma}n_{i\sigma^{\prime\prime}}\nonumber\\
& = & \hbar\omega_0\sum_i b{_i^{\dagger}}b_i-g\sum_{i,\sigma}\left(b{_i^{\dagger}}+b_i\right)n_{i\sigma}\nonumber\\
& + &\left(\frac{g^2}{\hbar\omega_0}\right)\sum_{i,\sigma}n_{i\sigma}n_{i\uparrow}
+\left(\frac{g^2}{\hbar\omega_0}\right)\sum_{i,\sigma}n_{i\sigma}n_{i\downarrow}\nonumber\\
& = &\hbar\omega_0\sum_i b{_i^{\dagger}}b_i-g\sum_{i,\sigma}\left(b{_i^{\dagger}}+b_i\right)n_{i\sigma}\nonumber\\
& + &\left(\frac{g^2}{\hbar\omega_0}\right)\sum_{i,\sigma}n{_{i\sigma}}
+\left(\frac{2g^2}{\hbar\omega_0}\right)\sum_{i}n_{i\uparrow}n_{i\downarrow}.
\label{eqn36}
\end{eqnarray}
In deriving Eq.~\ref{eqn36}, we use $\sigma^{\prime\prime}=\sigma$ or $\sigma^\prime$ ($\uparrow,\downarrow$). The last term of the 
ring Hamiltonian becomes
\begin{eqnarray}
g\sum_{i,\sigma}\left(\tilde{b}{_i^{\dagger}}+\tilde{b}_i\right)\tilde{n}_{i\sigma}& = & g\sum_{i,\sigma}\bigg[{b}{_i^{\dagger}}-
\left(\frac{g}{\hbar\omega_0}\right)\sum_{\sigma^{\prime\prime}} n_{i\sigma^{\prime\prime}}\nonumber\\
& + & b_i-\left(\frac{g}{\hbar\omega_0}\right)\sum_{\sigma^{\prime\prime}} n_{i\sigma^{\prime\prime}}\bigg]n_{i\sigma}\nonumber\\
& = &g\sum_{i,\sigma}\left(b{_i^{\dagger}}+b_i\right)n_{i\sigma}\nonumber\\
&-& \left(\frac{2g^2}{\hbar\omega_0}\right)\sum_{i,\sigma,\sigma{\prime\prime}}n_{i\sigma^{\prime\prime}}n_{i\sigma}\nonumber\\
& = & g\sum_{i,\sigma}\left(b{_i^{\dagger}}+b_i\right)n_{i\sigma}\nonumber\\
&-& \left(\frac{2g^2}{\hbar\omega_0}\right)\sum_{i,\sigma}n_{i\sigma}\nonumber\\
&-&\left(\frac{4g^2}{\hbar\omega_0}\right)\sum_{i}n_{i\uparrow}n_{i\downarrow}.
\label{eqn37}
\end{eqnarray}
For the above equation (Eq.~\ref{eqn37}), we use the conditions $n_{i\sigma}^2=n_{i\sigma}$ and $\left[n_{i\uparrow}\,,\,n_{i\downarrow}\right]=0$.

With all the above forms of the operators, the ring Hamiltonian (Eq.~\ref{eqn24}) can be expressed in a compact form as  
\begin{eqnarray}
\widetilde{H}_{R} & = & \sum_{i,\sigma}\epsilon_{i\sigma}c{_{i\sigma}^{\dagger}}c_{i\sigma}\nonumber\\
& + & \sum_{<i,j>,\sigma} \bigg\{c{_{i\sigma}^{\dagger}}c_{j\sigma}
\exp\left[\left(\frac{g}{\hbar\omega_0}\right)\left(b{_i^{\dagger}}-b_i\right)\right]\nonumber\\
&   &\exp\left[-\left(\frac{g}{\hbar\omega_0}\right)\left(b{_j^{\dagger}}-b_j\right)\right]+h.c\bigg\}\nonumber\\
& + & u \sum_{i,\sigma}n_{i\sigma}n_{i\sigma^{'}}+\hbar\omega_0\sum_i b{_i^{\dagger}}b_i-g\sum_{i,\sigma}\left(b{_i^{\dagger}}+b_i\right)n_{i\sigma}\nonumber\\ 
&+& \left(\frac{g^2}{\hbar\omega_0}\right)\sum_{i,\sigma}n_{i\sigma}+\left(\frac{2g^2}{\hbar\omega_0}\right)\sum_{i}
n_{i\uparrow}n_{i\downarrow}\nonumber\\
&+ &g\sum_{i,\sigma}\left(b{_i^{\dagger}}+b_i\right)n_{i\sigma}-\left(\frac{2g^2}{\hbar\omega_0}\right)\sum_{i,\sigma}n_{i\sigma}\nonumber\\
&-&\left(\frac{4g^2}{\hbar\omega_0}\right)\sum_{i}n_{i\uparrow}n_{i\downarrow}\nonumber\\
& = &\sum_{i,\sigma}\left(\epsilon_{i\sigma}-\frac{g^2}{\hbar\omega_0}\right)c{_{i\sigma}^{\dagger}}c_{i\sigma}\nonumber\\
& + &\sum_{<i,j>,\sigma} \bigg\{c{_{i\sigma}^{\dagger}}c_{j\sigma}
\exp\left[\left(\frac{g}{\hbar\omega_0}\right)\left(b{_i^{\dagger}}-b_i\right)\right]\nonumber\\
&   &\exp\left[-\left(\frac{g}{\hbar\omega_0}\right)\left(b{_j^{\dagger}}-b_j\right)\right]+h.c\bigg\}\nonumber\\
& + & \left(u-\frac{2g^2}{\hbar\omega_0}\right) \sum_{i}n_{i\uparrow}n_{i\downarrow}+\hbar\omega_0\sum_i 
b{_i^{\dagger}}b_i.
\label{eqn38}
\end{eqnarray}

\section{Zero-phonon averaging}
\label{ap2}

Here, we construct the effective ring Hamiltonian in pure electronic sub-space through zero-phonon averaging. We start with the operation
\begin{eqnarray}
H{_R^{eff}} & = & \langle\Phi_{ph}|\tilde{H}_{R}|\Phi_{ph}\rangle
\label{eqn39}
\end{eqnarray}
where, $\langle\Phi_{ph}|\Phi_{ph}\rangle = 1$.

At $\mathcal{T}=0\,$K, let us assume $|\Phi_{ph}\rangle = |0_{ph}\rangle$, where $|0_{ph}\rangle$ is the phonon vacuum state. Thus,
Eq.~\ref{eqn39} becomes
\begin{eqnarray}
H{_R^{eff}} & = & \langle 0_{ph}|\tilde{H}_{R}| 0_{ph}\rangle.
\label{eqn40}
\end{eqnarray}
With the zero-phonon averaging we get
\begin{eqnarray}
H{_R^{eff}} & = & \langle 0_{ph}|\widetilde{H}_R|0_{ph}\rangle \nonumber\\
& = & \sum_{i,\sigma}\left(\epsilon_{i\sigma}-\frac{g^2}{\hbar\omega_0}\right)c{_{i\sigma}^{\dagger}}c_{i\sigma}\,\,\langle 0_{ph}|0_{ph}\rangle\nonumber\\
& +&t\sum_{<i,j>,\sigma} \biggr\{c{_{i\sigma}^{\dagger}}c_{j\sigma}\,\,
\biggl< 0_{ph}\bigg|\,\,\exp\left[\left(\frac{g}{\hbar\omega_0}\right)\left(b{_i^{\dagger}}-b_i\right)\right]\nonumber\\
&   &\exp\left[-\left(\frac{g}{\hbar\omega_0}\right)\left(b{_j^{\dagger}}-b_j\right)\right]\bigg|0_{ph}\biggr> +h.c\biggr\}\nonumber\\
& + & \left(u-\frac{2g^2}{\hbar\omega_0}\right) \sum_{i}n_{i\uparrow}n_{i\downarrow}\,\,\langle 0_{ph}|0_{ph}\rangle\nonumber\\
& + & \hbar\omega_0\sum_i \,\,\langle 0_{ph}|b{_i^{\dagger}}b_i|0_{ph}\rangle\nonumber\\
& = & \sum_{i,\sigma}\left(\epsilon_{i\sigma}-\frac{g^2}{\hbar\omega_0}\right)c{_{i\sigma}^{\dagger}}c_{i\sigma}\nonumber\\
& +&t\sum_{<i,j>,\sigma} \biggr\{c{_{i\sigma}^{\dagger}}c_{j\sigma}\,\,
\biggl< 0_{ph}\bigg|\,\,\exp\left[\left(\frac{g}{\hbar\omega_0}\right)\left(b{_i^{\dagger}}-b_i\right)\right]\nonumber\\
&   &\exp\left[-\left(\frac{g}{\hbar\omega_0}\right)\left(b{_j^{\dagger}}-b_j\right)\right]\bigg|0_{ph}\biggr> +h.c\biggr\}\nonumber\\
& + & \left(u-\frac{2g^2}{\hbar\omega_0}\right) \sum_{i}n_{i\uparrow}n_{i\downarrow}\nonumber\\
& + & \hbar\omega_0\sum_i \,\,\langle 0_{ph}|b{_i^{\dagger}}b_i|0_{ph}\rangle. \nonumber\\
\label{eqn41}
\end{eqnarray}
The $2$nd term of Eq.~\ref{eqn41} still contains the phononic operators. We further simplify it through the following steps.
\begin{eqnarray}
& &\biggl< 0_{ph}\bigg|\,\,\exp\left[\left(\frac{g}{\hbar\omega_0}\right)\left(b{_i^{\dagger}}-b_i\right)\right]\nonumber\\
& &\exp\left[-\left(\frac{g}{\hbar\omega_0}\right)\left(b{_j^{\dagger}}-b_j\right)\right]\bigg|0_{ph}\biggr>\nonumber\\
& = & \biggl< 0_{ph}\bigg|\,\,\exp\left[\left(\frac{g}{\hbar\omega_0}\right)\left(b{_i^{\dagger}}-b_i\right)\right]\bigg|0_{ph}\biggr>\nonumber\\
& &\biggl< 0_{ph}\bigg|\,\,\exp\left[-\left(\frac{g}{\hbar\omega_0}\right)\left(b{_j^{\dagger}}-b_j\right)\right]\bigg|0_{ph}\biggr>.\nonumber\\
\label{eqn42}
\end{eqnarray}
Now, we use the relation
\begin{eqnarray}
e^{A+B} & = & e^{-\frac{1}{2}\left[A\,,\,B\right]}\,e^A\,e^B. \nonumber\\
\mbox{If},~ A & = & \left(\frac{g}{\hbar\omega_0}\right)b{_i^{\dagger}}, \nonumber\\
\mbox{and}~   B & = &-\left(\frac{g}{\hbar\omega_0}\right)b_i, \nonumber
\end{eqnarray}
then,
\begin{eqnarray}
\left[A\,,\,B\right] & = & -\left(\frac{g}{\hbar\omega_0}\right)^2\left[b{_i^{\dagger}}\,,\,b_i\right]\nonumber\\
& = & \left(\frac{g}{\hbar\omega_0}\right)^2.
\end{eqnarray}
Following these expressions, we get   
\begin{eqnarray}
& &\biggl< 0_{ph}\bigg|\exp\left[\left(\frac{g}{\hbar\omega_0}\right)\left(b{_i^{\dagger}}-b_i\right)\right]\bigg|0_{ph}\biggr>\nonumber\\
& = & \exp\left[-\frac{1}{2}\left(\frac{g}{\hbar\omega_0}\right)^2\right]\biggl< 0_{ph}\bigg|\,\,\exp\left[\left(\frac{g}{\hbar\omega_0}\right)b{_i^{\dagger}}\right]\nonumber\\
& &\exp\left[-\left(\frac{g}{\hbar\omega_0}\right)b_i\right]\bigg|0_{ph}\biggr>.
\label{eqn43}
\end{eqnarray}
Now, operating the phononic creation and annihilation operators on the phononic $n$th state, it can be found
\begin{equation}
\langle n|e^{z^{*}b^{\dagger}}e^{-zb}|n \rangle = L_n(|z|^2)\nonumber
\end{equation}
where, $L_n$ is the $n$th Lagurre polynomial of order $n$. Considering the vacuum state, we have
\begin{eqnarray}
& &\biggl< 0_{ph}\bigg|\exp\left[\left(\frac{g}{\hbar\omega_0}\right)b{_i^{\dagger}}\right]\exp\left[-\left(\frac{g}{\hbar\omega_0}\right)b_i\right]\bigg|0_{ph}\biggr>\nonumber\\
& = &L_0\left(\bigg|\frac{g}{\hbar\omega_0}\bigg|^2\right)=1.
\label{eqn44}
\end{eqnarray}
Similarly,
\begin{eqnarray}
& &\biggl< 0_{ph}\bigg|\exp\left[-\left(\frac{g}{\hbar\omega_0}\right)b{_j^{\dagger}}\right]\exp\left[\left(\frac{g}{\hbar\omega_0}\right)b_j\right]\bigg|0_{ph}\biggr>\nonumber\\
& = &\biggl< 0_{ph}\bigg|\exp\left[\left(\frac{g}{\hbar\omega_0}\right)b{_j^{\dagger}}\right]\exp\left[-\left(\frac{g}{\hbar\omega_0}\right)
b_j\right]\bigg|0_{ph}\biggr>^{\dagger}\nonumber\\
& = & 1.
\end{eqnarray}
Now, the terms of Eq.~\ref{eqn42} can be expressed as 
\begin{eqnarray}
& &\biggl< 0_{ph}\bigg|\exp\left[\left(\frac{g}{\hbar\omega_0}\right)\left(b{_i^{\dagger}}-b_i\right)\right]\bigg|0_{ph}\biggr>\nonumber\\
& = & \biggl< 0_{ph}\bigg|\exp\left[-\left(\frac{g}{\hbar\omega_0}\right)\left(b{_j^{\dagger}}-b_j\right)\right]\bigg|0_{ph}\biggr>\nonumber\\
& = & \exp\left[-\frac{1}{2}\left(\frac{g}{\hbar\omega_0}\right)^2\right].
\end{eqnarray}
Hence, Eq.~\ref{eqn42} reads as
\begin{eqnarray}
& &\biggl< 0_{ph}\bigg|\,\,\exp\left[\left(\frac{g}{\hbar\omega_0}\right)\left(b{_i^{\dagger}}-b_i\right)\right]\nonumber\\
& &\exp\left[-\left(\frac{g}{\hbar\omega_0}\right)\left(b{_j^{\dagger}}-b_j\right)\right]\bigg|0_{ph}\biggr>\nonumber\\
& = &\exp\left[-\frac{1}{2}\left(\frac{g}{\hbar\omega_0}\right)^2\right]\exp\left[-\frac{1}{2}\left(\frac{g}{\hbar\omega_0}\right)^2\right]\nonumber\\
& = &\exp\left[-\left(\frac{g}{\hbar\omega_0}\right)^2\right].
\label{eqn45}
\end{eqnarray}
The last term of Eq.~\ref{eqn41} is
\begin{equation}
\langle 0_{ph}|b{_i^{\dagger}}b_i|0_{ph}\rangle=0.
\label{eqn46}
\end{equation}
Substituting Eq.~\ref{eqn45} and Eq.~\ref{eqn46} in Eq.~\ref{eqn41}, we finally get the effective ring Hamiltonian in the electronic 
sub-space as 
\begin{eqnarray}
H{_R^{eff}} & = & \sum_{i,\sigma}\left(\epsilon_{i\sigma}-\frac{g^2}{\hbar\omega_0}\right)c{_{i\sigma}^{\dagger}}c_{i\sigma}\nonumber\\
& +&t\exp\left[-\left(\frac{g}{\hbar\omega_0}\right)^2\right]\sum_{<i,j>,\sigma} \biggr\{c{_{i\sigma}^{\dagger}}c_{j\sigma}+h.c\biggr\}\nonumber\\
& + & \left(u-\frac{2g^2}{\hbar\omega_0}\right) \sum_{i}n_{i\uparrow}n_{i\downarrow}.
\label{eqn47}
\end{eqnarray}

\section{Modification of the coupling Hamiltonian in the presence of e-ph coupling}
\label{ap3}

Here, we derive the modified coupling Hamiltonian due to the e-ph coupling in the ring system. We start with the coupling Hamiltonian
\begin{eqnarray}
H_{coupling}=\sum_\sigma\left(\tau_S c{_{1\sigma}^{\dagger}} d_{0\sigma} + \tau_D c{_{p\sigma}^{\dagger}} d_{N+1\sigma}+h.c.\right).
\label{eqn48}
\end{eqnarray}
Similar to the earlier prescription mentioned in Appendix~\ref{ap1}, we get
\begin{eqnarray}
\widetilde{H}_{coupling} & = & A\,H_{coupling}\,A^{-1} \nonumber \\
& = &\sum_\sigma A \left(\tau_S c{_{1\sigma}^{\dagger}} d_{0\sigma}+\tau_D c{_{p\sigma}^{\dagger}} d_{N+1\sigma}+h.c.\right)A^{-1} \nonumber\\
& = & \sum_\sigma \left(\tau_S Ac{_{1\sigma}^{\dagger}}A^{-1}Ad_{0\sigma}A^{-1}\right.\nonumber \\ 
& & \left. +\tau_D A c{_{p\sigma}^{\dagger}}A^{-1}Ad_{N+1\sigma}A^{-1}+h.c.\right)\nonumber\\
& = &\sum_\sigma \left(\tau_S\tilde{c}{_{1\sigma}^{\dagger}} \tilde{d}_{0\sigma}+\tau_D\tilde{c}{_{p\sigma}^{\dagger}} \tilde{d}_{N+1\sigma}+h.c.\right).
\label{eqn49}
\end{eqnarray}
Here, the leads are considered to be free from any interactions and the Lang-Firsov operator $U$ only contains the fermionic and 
bosonic operators associated with the ring conductor only. Consequently, $U$ commutes with the fermionic operators of the leads.
Hence,
\begin{eqnarray}
\tilde{d}_{0\sigma}= e^U\,d_{0\sigma}\, e^{-U} = d_{0\sigma}, \nonumber 
\end{eqnarray}
and
\begin{eqnarray}
\tilde{d}_{N+1\sigma}= e^U\,d_{N+1\sigma}\,e^{-U}= d_{N+1\sigma}. \nonumber
\end{eqnarray}
Then, the coupling Hamiltonian reads as,
\begin{eqnarray}
\widetilde{H}_{coupling} & = & \sum_\sigma \bigg\{\tau_S c{_{1\sigma}^{\dagger}}d_{0\sigma}\exp\left[\left(\frac{g}{\hbar\omega_0}\right)
\left(b{_1^{\dagger}}-b_1\right)\right]\nonumber\\
&+&\tau_D c{_{p\sigma}^{\dagger}}d_{N+1\sigma}\exp\left[\left(\frac{g}{\hbar\omega_0}\right)\left(b{_p^{\dagger}}-b_p\right)\right]\nonumber\\
& + & h.c.\bigg\}.
\end{eqnarray}
Now, we need to do the zero-phonon averaging to get the effective coupling Hamiltonian following the same method as prescribed 
in Appendix~\ref{ap2}. The effective coupling Hamiltonian becomes
\begin{eqnarray}
H{_{coupling}^{eff}} & = & \langle0_{ph}|\widetilde{H}_{coupling}|0_{ph}\rangle\nonumber\\
& = & \sum_\sigma \bigg\{\tau_S c{_{1\sigma}^{\dagger}}d_{0\sigma}\nonumber\\
&&\bigg\langle0_{ph}\bigg|\exp\left[\left(\frac{g}{\hbar\omega_0}\right)\left(b{_1^{\dagger}}-b_1\right)\right]\bigg|0_{ph}\bigg\rangle
\nonumber\\
&+&\tau_D c{_{p\sigma}^{\dagger}}d_{N+1\sigma}\nonumber\\
&&\bigg\langle0_{ph}\bigg|\exp\left[\left(\frac{g}{\hbar\omega_0}\right)\left(b{_p^{\dagger}}-b_p\right)\right]\bigg|0_{ph}\bigg\rangle
 + h.c. \bigg\}  \nonumber\\
& = & \sum_\sigma e^{-\frac{1}{2}(\frac{g}{\hbar \omega_0})^2}\left(c{_{1\sigma}^{\dagger}}d_{0\sigma}+c{_{p\sigma}^{\dagger}}d_{N+1\sigma}+h.c.\right) \nonumber\\
&=& e^{-\frac{1}{2}(\frac{g}{\hbar \omega_0})^2} H_{coupling}.
\end{eqnarray}

\end{document}